\documentclass[12pt]{article}
\usepackage{graphicx}
\usepackage{epsfig}
\usepackage{epstopdf}
\DeclareGraphicsExtensions{.pdf,.eps,.png,.jpg,.mps}
\usepackage{caption}
\usepackage{float}

\usepackage{cite}

\def\lsim{\mathrel{\rlap{\lower3pt\hbox{\hskip0pt$\sim$}}
     \raise1pt\hbox{$<$}}}         
\def\gsim{\mathrel{\rlap{\lower4pt\hbox{\hskip1pt$\sim$}}
     \raise1pt\hbox{$>$}}}         

\usepackage{amsmath}
\usepackage{amsfonts}

\begin{document}
\begin{titlepage}

\centerline{\Large \bf Statistical Industry Classification}
\medskip

\centerline{Zura Kakushadze$^\S$$^\dag$\footnote{\, Zura Kakushadze, Ph.D., is the President of Quantigic$^\circledR$ Solutions LLC,
and a Full Professor at Free University of Tbilisi. Email: zura@quantigic.com} and Willie Yu$^\sharp$\footnote{\, Willie Yu, Ph.D., is a Research Fellow at Duke-NUS Medical School. Email: willie.yu@duke-nus.edu.sg}}
\bigskip

\centerline{\em $^\S$ Quantigic$^\circledR$ Solutions LLC}
\centerline{\em 1127 High Ridge Road \#135, Stamford, CT 06905\,\,\footnote{\, DISCLAIMER: This address is used by the corresponding author for no
purpose other than to indicate his professional affiliation as is customary in
publications. In particular, the contents of this paper
are not intended as an investment, legal, tax or any other such advice,
and in no way represent views of Quantigic$^\circledR$ Solutions LLC,
the website \underline{www.quantigic.com} or any of their other affiliates.
}}
\centerline{\em $^\dag$ Free University of Tbilisi, Business School \& School of Physics}
\centerline{\em 240, David Agmashenebeli Alley, Tbilisi, 0159, Georgia}
\centerline{\em $^\sharp$ Centre for Computational Biology, Duke-NUS Medical School}
\centerline{\em 8 College Road, Singapore 169857}
\medskip
\centerline{(June 29, 2016)}

\bigskip
\medskip

\begin{abstract}
{}We give complete algorithms and source code for constructing (multilevel) statistical
industry classifications, including methods for fixing the number of clusters at each level (and the number of levels).
Under the hood there are clustering algorithms (e.g., k-means). However, what should we cluster?
Correlations? Returns? The answer turns out to be neither and our backtests suggest that these details
make a sizable difference. We also give an algorithm and source code for building ``hybrid" industry classifications
by improving off-the-shelf ``fundamental" industry classifications by applying our statistical industry classification methods to them.
The presentation is intended to be pedagogical and geared toward practical applications in quantitative trading.
\end{abstract}
\medskip
\end{titlepage}

\newpage

\section{Introduction and Summary}

{}Industry classifications such as GICS, BICS, ICB, NAICS, SIC, etc.\footnote{\, Hereinafter we will refer to these as ``fundamental" industry classifications (see below).} are widely used in quantitative trading. They group stocks into baskets, e.g., industries, i.e., based on some kind of a similarity criterion. On general grounds one then expects (or hopes) that stocks within such baskets on average should be relatively highly correlated. This is valuable information and can be used in various ways. E.g., one can build a simple mean-reversion statistical arbitrage strategy whereby one assumes that stocks in a given industry move together, cross-sectionally demeans stock returns within said industry, shorts stocks with positive residual returns and goes long stocks with negative residual returns, with some generally nonuniform weights.\footnote{\, More generally, one employs a weighted regression instead of demeaning, and there are various ways of fixing the aforesaid weights. For a pedagogical discussion, see, e.g., \cite{MeanRev}.} Industries can also be used as risk factors in multifactor risk models.\footnote{\, For a discussion and literature on multifactor risk models, see, e.g., \cite{GrinoldKahn}.}

{}The aforementioned ``fundamental" industry classifications are based on grouping companies together based on fundamental/economic data (see Section \ref{sec2}), which is expected to add value on longer holding horizons. What about shorter holding horizons relevant to quantitative trading strategies? Other than a large number of market players using such industry classifications to arbitrage mispricings,\footnote{\, This very relevant reason should not to be underestimated, despite its ``behavioral" nature.} how do we know that they are competitive with purely statistical methods at short horizons?

{}It is no secret that modern quantitative trading heavily relies on statistical methods such as data mining, machine learning, clustering algorithms, etc. However, after all, quantitative trading is a secretive field and resources on how things are done in practice are at best scarce.\footnote{\, Thus, we are unaware of another paper discussing the material herein at short horizons.} The purpose of these notes is to discuss a systematic quantitative framework -- in what is intended to be a ``pedagogical" fashion -- for building what we refer to as {\em statistical industry classifications}, solely based on stock returns and no additional extraneous data. Under the hood we have clustering algorithms. However, picking a clustering algorithm -- and we will see that some work better than others -- is insufficient. E.g., what should we cluster? Correlations? Returns? The answer turns out to be neither and stems from quantitative trading intuition, which is not something one expects to find in machine learning books. We discuss various nuances in constructing statistical industry classifications, and it is those nuances that make a sizable difference. Quant trading is all about detail.

{}One motivation for considering statistical industry classifications -- apart from the evident, to wit, the fact that they differ from ``fundamental" industry classifications and are widely used in quant trading -- is scenarios where ``fundamental" industry classifications are unavailable (or are of subpar quality). This could be in emerging or smaller markets, or even in the U.S. if the underlying trading portfolios are relatively small and a ``fundamental" industry classification produces too fragmented a grouping. However, perhaps an equally -- if not more -- important motivation is application of these methods to returns for ``instruments" other than stocks, e.g., quantitative trading alphas, for which there is no analog of a ``fundamental" industry classification \cite{Billion}. We will keep this in mind below.\footnote{\, Optimizing weights in alpha portfolios has its own nuances \cite{Billion}; however, the methods we discuss here are readily portable to alpha returns as they are purely statistical. Here we backtests them (see below) on stock returns as the historical data is readily available. Alpha return time series are highly proprietary, so publishing backtests is not feasible.}

{}In Section \ref{sec2} we briefly review some generalities of (binary) ``fundamental" industry classifications to set up the framework for further discussion. Next, in Section \ref{sec.stat.clust} we address the issue of what to cluster. We discuss why clustering correlations is suboptimal, and why so is directly clustering returns. We argue that returns should be normalized before clustering and give an explicit prescription for such normalization. We then discuss how to construct single-level and multilevel (hierarchical -- e.g., BICS has 3 levels: sectors, industries and sub-industries) statistical industry classifications together with some tweaks (e.g., cross-sectionally demeaning returns at less granular levels). Many clustering algorithms such as k-means are not deterministic. This can be a nuisance. We give an explicit prescription for aggregating classifications from multiple samplings, which in fact improves stability and performance. We discuss algorithms for ``bottom-up" (most granular to least granular level), ``top-down" (least granular to most granular level) and ``relaxation" (hierarchical agglomerative) clustering, together with their ``pros" and ``cons".

{}In Section \ref{sec.backtests} we discuss detailed backtests of the various algorithms in Section \ref{sec.stat.clust} and subsequent sections utilizing the intraday alphas and backtesting procedure described in \cite{Het} by using the resultant multilevel statistical industry classifications for building heterotic risk models. The backtests unequivocally suggest that there is structure in the return time series beyond what is captured by simple principal component analysis and clustering adds value. However, clustering still cannot compete with ``fundamental" industry classifications in terms of performance due to inherent out-of-sample instabilities in any purely statistical algorithm.

{}In Section \ref{sec.dyn} we take it a step further and give a prescription for fixing the number of clusters at each level using the methods discussed in \cite{StatRM}, including eRank (effective rank) defined in \cite{RV}. We also discuss a heuristic for fixing the number of levels, albeit we empirically observe that the number of levels is not as influential as the number of clusters, at least in our backtests. We take this even further in Section \ref{sec.hybrid}, where we give an algorithm for improving a ``fundamental" industry classification via further clustering large sub-industries (using BICS nomenclature) at the most granular level via statistical industry classification algorithms we discuss here thereby increasing granularity and improving performance. We briefly conclude in Section \ref{sec.concl} and outline some ideas.

{}We give the R source code for our algorithms in Appendix \ref{app.A} (multilevel ``bottom-up" clustering, dynamical cluster numbers), Appendix \ref{app.B} (multilevel ``top-down" clustering) and Appendix \ref{app.C} (``relaxation" clustering). Appendix \ref{app.D} contains legalese.

\section{Industry Classification}\label{sec2}

{}An industry classification is based on a similarity criterion: stocks' membership in ``groups" or ``clusters" such as sectors, industries, sub-industries, etc. -- the nomenclature varies from one industry classification scheme to another. Commonly used industry classifications such as GICS, BICS, ICB, NAICS, SIC, etc., are based on fundamental/economic data (such as companies' products and services and more generally their revenue sources, suppliers, competitors, partners, etc.). Such industry classifications are essentially independent of the pricing data and, if well-built, tend to be rather stable out-of-sample as companies seldom jump industries.\footnote{\, However, there is variability in the performance of different industry classifications.}

{}An industry classification can consist of a single level: $N$ tickers labeled by $i=1,\dots,N$ are grouped into $K$ ``groups" -- let us generically call them ``clusters" -- labeled by $A=1,\dots,K$. So, we have a map $G:\{1,\dots,N\}\mapsto\{1,\dots,K\}$ between stocks and ``clusters".\footnote{\, Here we are assuming that each stock belongs to one and only one ``cluster". Generally, this assumption can be relaxed thereby allowing for ``conglomerates" that belong to multiple sub-industries, industries, sectors, etc. However, this is not required for our purposes here.} More generally, we can have a hierarchy with multiple levels. We can schematically represent this via: Stocks $\rightarrow$ Level-1 ``Clusters" $\rightarrow$ Level-2 ``Clusters" $\rightarrow \dots \rightarrow$ Level-$P$ ``Clusters". Let us label these $P$ levels by $\mu = 1,\dots, P$. Level-1 is the most granular level with $N$ stocks grouped into $K_1$ ``clusters". The Level-1 ``clusters" are in turn grouped into $K_2$ Level-2 ``clusters", where $K_2 < K_1$, and so on, Level-$P$ being least granular.\footnote{\, The branches in this hierarchy tree are assumed to have equal lengths. More generally, we can have branches of nonuniform lengths. However, shorter branches can always be extended to the length of the longest branch(es) by allowing single-element (including single-stock) ``clusters".} Thus, consider
BICS\footnote{\, Bloomberg Industry Classification System.} as an illustrative example, which has a 3-level hierarchy: Stocks $\rightarrow$ Sub-industries $\rightarrow$ Industries $\rightarrow$ Sectors. (Here ``Sub-industries" is the most granular level, while ``Sectors" is the least granular level.) So, we have: $N$ stocks labeled by $i=1,\dots,N$; $K$ sub-industries labeled by $A=1,\dots,K$; $F$ industries labeled by $a=1,\dots,F$; and $L$ sectors labeled by $\alpha=1,\dots,L$. Let $G$ be the map between stocks and sub-industries, $S$ be the map between sub-industries and industries, and $W$ be the map between industries and sectors:
\begin{eqnarray}\label{G.map}
 &&G:\{1,\dots,N\}\mapsto\{1,\dots,K\}\\
 &&S:\{1,\dots,K\}\mapsto\{1,\dots,F\}\label{S.map}\\
 &&W:\{1,\dots,F\}\mapsto\{1,\dots,L\}\label{W.map}
\end{eqnarray}
The beauty of such ``binary" industry classifications (generally, with $P$ levels) is that the ``clusters" (in the case of BICS, sub-industries, industries and sectors) can be used to identify blocks (sub-matrices) in the sample correlation matrix $\Psi_{ij}$ of stock returns.\footnote{\, And this is useful in constructing risk models for portfolio optimization \cite{Het}.} E.g., for sub-industries the binary matrix $\delta_{G(i), A}$ defines such blocks.

\section{Statistical Clustering}\label{sec.stat.clust}

{}What if we do not have access to industry classifications based on fundamental data\footnote{\, Commercially available industry classifications such as GICS and ICB come at nontrivial cost. The underlying SIC data is available from SEC for free, albeit only by company names, not by ticker symbols. It takes considerable effort to download this data and transform it into an actual industry classification. Alternatively, it can be purchased from commercial providers.} or one is unavailable for the stock universe we wish to trade? Can we build an industry classification from pricing data, i.e., directly from stock returns? After all, intuitively, the time series of returns contains information about how correlated the stocks are. Can we extract it and transform it into an industry classification?

{}The answer is yes, but it is tricky. The key issue is that correlations between stocks typically are highly unstable out-of-sample. A naive attempt at constructing an industry classification based on stock returns may produce an industry classification with subpar performance. Our goal here is to discuss how to mitigate the out-of-sample instability by building statistical industry classifications based on clustering quantities other than returns. But first let us discuss clustering itself.

\subsection{K-means}

{}A popular clustering algorithm is k-means \cite{Steinhaus}, \cite{Lloyd1957}, \cite{Forgy}, \cite{MacQueen}, \cite{Hartigan}, \cite{HartWong}, \cite{Lloyd1982}. The basic idea behind k-means is to partition $N$ observations into $K$ clusters such that each observation belongs to the cluster with the nearest mean. Each of the $N$ observations is actually a $d$-vector, so we have an $N \times d$ matrix $X_{is}$, $i=1,\dots,N$, $s=1,\dots,d$. Let $C_a$ be the $K$ clusters, $C_a = \{i| i\in C_a\}$, $a=1,\dots,K$. Then k-means attempts to minimize
\begin{equation}\label{k-means}
 g = \sum_{a=1}^K \sum_{i \in C_a} \sum_{s=1}^d  \left(X_{is} - Y_{as}\right)^2
\end{equation}
where
\begin{equation}\label{centers}
 Y_{as} = {1\over n_a} \sum_{i\in C_a} X_{is}
\end{equation}
are the cluster centers (i.e., cross-sectional means),\footnote{\, Throughout this paper ``cross-sectional" refers to ``over the index $i$".} and $n_a = |C_a|$ is the number of elements in the cluster $C_a$. In (\ref{k-means}) the measure of ``closeness" is chosen to be the Euclidean distance between points in ${\bf R}^d$, albeit other measures are possible.

{}One ``drawback" of k-means is that it is not a deterministic algorithm. Generically, there are copious local minima of $g$ in (\ref{k-means}) and the algorithm only guarantees that it will converge to a local minimum, not the global one. Being an iterative algorithm, k-means starts with a random or user-defined set of the centers $Y_{as}$ at the initial iteration. However, as we will see, this ``drawback" actually adds value.

\subsection{What to Cluster?}

{}So, what should we cluster to construct statistical industry classifications? I.e., what should we pick as our matrix $X_{is}$ in (\ref{k-means})? It is tempting to somehow use pair-wise stock correlations. However, the sample correlation matrix $\Psi_{ij}$ computed based on the time series of stock returns is highly unstable out-of-sample.\footnote{\, The sample correlation matrix contains less information than the underlying time series of returns. Thus, it knows nothing about serial means of returns, only deviations from these means.} So, what if we identify $X_{is}$ with the time series of the underlying stock returns? Let $R_{is}$ be these stock returns, where $s = 1,\dots,d$ now is interpreted as labeling the observations in the time series (e.g., trading days). Further, for definiteness, let $s=1$ correspond to the most recent observation. Now we can build a statistical industry classification by applying k-means to $X_{is} = R_{is}$. Intuitively this makes sense: we are clustering stocks based on how close the returns are to the centers (i.e., within-cluster cross-sectional means) of the clusters they belong to. However, this is a suboptimal choice.

{}Indeed, this can be understood by observing that, in the context of stock returns, a priori there is no reason why the centers $Y_{as}$ in (\ref{centers}) should be computed with equal weights. We can think of the clusters $C_a$ as portfolios of stocks, and $Y_{as}$ as the returns for these portfolios. Therefore, based on financial intuition, we may wish to construct these portfolios with nonuniform weights. Furthermore, upon further reflection, it become evident that clustering returns make less sense than it might have appeared at first. Indeed, stock volatility is highly variable, and its cross-sectional distribution is not even quasi-normal but highly skewed, with a long tail at the higher end -- it is roughly log-normal. Clustering returns does not take this skewness into account and inadvertently we might be clustering together returns that are not at all highly correlated solely due to the skewed volatility factor.

{}A simple solution is to cluster the normalized returns ${\widetilde R}_{is} = R_{is} / \sigma_i$, where $\sigma_i^2 = \mbox{Var}(R_{is})$ is the serial variance. This way we factor out the skewed volatility factor. Indeed, $\mbox{Cov}({\widetilde R}_i, {\widetilde R}_j) = \mbox{Cor}(R_i, R_j) = \Psi_{ij}$ (we suppress the index $s$ in the serial covariance Cov and correlation Cor) is the sample correlation matrix with $|\Psi_{ij}| \leq 1$. However, as we will see below, clustering ${\widetilde R}_{is}$, while producing better results than clustering $R_{is}$, is also suboptimal. Here are two simple arguments why this is so.

{}Clusters $C_a$ define $K$ portfolios whose weights are determined by what we cluster. When we cluster $X_{is} = R_{is}$, the centers are $Y_{as} = \mbox{Mean}(R_{is}|i\in C_a)$, i.e., we have equal weights $\omega_i\equiv 1$ for the aforesaid $K$ portfolios, and we group $R_{is}$ (at each iterative step in the k-means algorithm) by how close these returns are to these equally-weighted portfolios. However, equally-weighted portfolios themselves are suboptimal. So are portfolios weighted by $\omega_i\equiv 1/\sigma_i$, which is what we get if we cluster $X_{is} = {\widetilde R}_{is}$, where the centers are $Y_{as} = \mbox{Mean}(R_{is}/\sigma_i|i\in C_a)$. Thus, portfolios that maximize the Sharpe ratio \cite{Sharpe1994} are weighted by inverse variances:\footnote{\, More precisely, this is the case in the approximation where the sample covariance matrix is taken to be diagonal. In the context of clustering it makes sense to take the diagonal part of the sample covariance matrix as the full sample covariance matrix is singular for clusters with $n_a > d-1$. Even for $n_a \leq d-1$ the sample covariance matrix, while invertible, has highly out-of-sample unstable off-diagonal elements. In contrast, the diagonal elements, i.e., sample variances $\sigma_i^2$, are much more stable, even for short lookbacks. So it makes sense to use them in defining $\omega_i$.} $\omega_i=1/\sigma_i^2$. We get such portfolios if we cluster $X_{is} = {\widehat R}_{is}$, where ${\widehat R}_{is} = R_{is}/\sigma_i^2$, so the centers are $Y_{as} = \mbox{Mean}(R_{is}/\sigma_i^2|i\in C_a)$. Clustering ${\widehat R}_{is}$, as we will see, indeed outperforms clustering ${\widetilde R}_{is}$. Can we understand this in a simple, intuitive fashion?

{}By clustering ${\widetilde R}_{is} = R_{is}/\sigma_i$, we already factor out the volatility dependence. So, why would clustering ${\widehat R}_{is} = R_{is}/\sigma_i^2$ work better? Clustering ${\widetilde R}_{is}$ essentially groups together stocks that are (to varying degrees) highly correlated in-sample. However, there is no guarantee that they will remain as highly correlated out-of-sample. Intuitively, it is evident that higher volatility stocks are more likely to get uncorrelated with their respective clusters. This is essentially why suppressing by another factor or $\sigma_i$ in ${\widehat R}_{is}$ (as compared with ${\widetilde R}_{is}$) leads to better performance: inter alia, it suppresses contributions of those volatile stocks into the cluster centers $Y_{is}$.

\subsubsection{A Minor Tweak}\label{sub.norm.ret}

{}So, we wish to cluster ${\widehat R}_{is} = R_{is}/\sigma_i^2$. There is a potential hiccup with this in practice. If some stocks have very low volatilities, we could have large ${\widehat R}_{is}$ for such stocks. To avoid any potential issues with computations, we can ``smooth" this out via (MAD = mean absolute deviation):\footnote{\, This is one possible tweak. Others produce similar results.}
\begin{eqnarray}\label{tweak}
 &&{\widehat R}_{is} = {R_{is} \over {\sigma_i u_i}}\\
 &&u_i = {\sigma_i\over v}\\
 &&v = \exp(\mbox{Median}(\ln(\sigma_i)) - 3~\mbox{MAD}(\ln(\sigma_i)))
\end{eqnarray}
and for all $u_i < 1$ we set $u_i \equiv 1$. This is the definition of ${\widehat R}_{is}$ we use below (unless stated otherwise). Furthermore, Median($\cdot$) and MAD($\cdot$) above are cross-sectional.

\subsection{Multilevel Clustering}

{}If we wish to construct a single-level statistical industry classification, we can simply cluster ${\widehat R}_{is}$ defined in (\ref{tweak}) into $K$ clusters via k-means. What if we wish to construct a multilevel statistical industry classification (see Section \ref{sec2})? We discuss two approaches here, which we can refer to as ``bottom-up" and ``top-down".\footnote{\, W.r.t. classification levels; ``bottom-up" should not be confused with agglomerative clustering.}

\subsubsection{Bottom-Up Clustering}\label{sub.bottomup}

{}Say we wish to construct a $P$-level classification. We can construct it as a sequence: $K_1 \rightarrow K_2 \rightarrow \dots \rightarrow K_P$ ($K_1 > K_2 > \dots > K_P$), where we first construct the most granular level with $K_1$ clusters, then we cluster these $K_1$ clusters into fewer $K_2$ clusters and so on, until we reach the last and least granular level with $K_P$ clusters. Given\footnote{\, We will discuss what these cluster number ``should" be below.} the integers $K_1,\dots,K_P$, the question is what to use as the returns at each step. Let these returns be $[R(\mu)]_{i(\mu), s}$ (i.e., we cluster $[R(\mu)]_{i(\mu), s}$ into $K_\mu$ clusters via k-means), where $\mu =1,\dots,P$, $i(\mu) = 1,\dots,K_{\mu-1}$, and we have conveniently defined $K_0 = N$, so $i(1)$ is the same index as $i$. As above, we can take $[R(1)]_{is} = {\widehat R}_{is}$. What about   $[R(\mu)]_{i(\mu), s}$ at higher levels $\mu > 1$? We have some choices here. Let $C_{a(\mu)} = \{i(\mu)| i(\mu)\in C_{a(\mu)}\}$, $a(\mu)=1,\dots,K_\mu$ be the clusters at each level $\mu$. I.e., the index $a(\mu)$ is the same as the index $i(\mu+1)$ for $0 < \mu < P$. Then we can take (in the second line below $2 < \mu \leq P$)
\begin{eqnarray}
 &&[R(2)]_{i(2), s} = \mbox{Mean}(R^\prime_{is}| i \in \{1,\dots,N\})\\
 &&[R(\mu)]_{i(\mu), s} = \mbox{Mean}([R^\prime(\mu-1)]_{i(\mu-1),s}| i(\mu-1) \in C_{a(\mu - 1)})\label{Rmu}
\end{eqnarray}
where we can take (i) $R^\prime_{is} = R_{is}$ and $[R^\prime(\mu)]_{i(\mu),s} = [R(\mu)]_{i(\mu),s}$, or (ii) $R^\prime_{is} = {\widehat R}_{is}$ and $[R^\prime(\mu)]_{i(\mu),s} = [{\widehat R}(\mu)]_{i(\mu),s}$, where (Var($\cdot$) below is the serial variance)
\begin{eqnarray}
 &&[{\widehat R}(\mu)]_{i(\mu),s} = {[R(\mu)]_{i(\mu),s} \over \sigma_{i(\mu)}^2}\\
 &&\sigma_{i(\mu)}^2 = \mbox{Var}([R(\mu)]_{i(\mu),s})
\end{eqnarray}
These two definitions produce very similar results in our backtests (see below).

\subsubsection{Another Minor Tweak}\label{sub.demean}

{}In the bottom-up clustering approach we just discussed above, the higher level clusters tend to be highly correlated with each other. I.e., the corresponding cluster returns have a prominent ``market" (or ``overall") mode\footnote{\, See, e.g., \cite{CFM}, \cite{Billion}.} component in them. That is, averages of pair-wise ($i(\mu)\neq j(\mu)$) serial correlations $[\Psi(\mu)]_{i(\mu),j(\mu)} = \mbox{Cor}([R(\mu)]_{i(\mu), s}, [R(\mu)]_{j(\mu), s})$ at higher levels $\mu > 1$ are substantial.\footnote{\, Consequently, there is a large gap between the first $[\lambda(\mu)]^{(1)}$ and higher $[\lambda(\mu)]^{(p)}$, $p>1$, eigenvalues of $[\Psi(\mu)]_{i(\mu),j(\mu)}$; the eigenvalues are ordered decreasingly: $[\lambda(\mu)]^{(1)} > [\lambda(\mu)]^{(2)} > \dots$} To circumvent this, we can simply cross-sectionally demean the returns at higher levels, i.e., for $\mu > 1$ we substitute $[R(\mu)]_{i(\mu), s}$ by $[R(\mu)]_{i(\mu), s} - \mbox{Mean}([R(\mu)]_{i(\mu), s} | i(\mu)\in C_a(\mu))$. However, cross-sectional demeaning at level-1 ($\mu = 1$) leads to worse performance. Intuitively, we can understand this as follows. Demeaning at the most granular level removes the ``market" mode.\footnote{\, This essentially drops the 1st principal component from the spectral decomposition of $\Psi_{ij}$.} Unlike higher-level returns $[R(\mu)]_{i(\mu), s}$, $\mu > 1$, the level-1 returns are not all that highly correlated with each other, so it pays to keep the ``market" mode intact as, e.g., high-beta stocks statistically are expected to cluster together, while low-beta stocks are expected to cluster differently. So, the upshot is that we demean the returns at higher levels, but not level-1 returns.

\subsubsection{Aggregating Multiple Samplings}\label{sub.aggr}

{}As mentioned above, k-means is not a deterministic algorithm. Unless the initial centers are preset, the algorithm starts with random initial centers and converges to a different local minimum in each run. There is no magic bullet here: trying to ``guess" the initial centers is not any easier than ``guessing" where, e.g., the global minimum is. So, what is one to do? One possibility is to simply live with the fact that every run produces a different answer. The question then one must address in a given context is whether the performance in an actual application is stable from one such random run to another, or if it is all over the place. As we will see below, in our backtests, happily, the performance is extremely stable notwithstanding the fact that each time k-means produces a different looking industry classification.

{}So, this could be the end of the story here. However, one can do better. The idea is simple. What if we {\em aggregate} different industry classifications from multiple runs (or samplings) into one? The question is how. Suppose we have $M$ runs ($M \gg 1$). Each run produces an industry classification with $K$ clusters. Let $\Omega^r_{ia} = \delta_{G^r(i),a}$, $i=1,\dots,N$, $a=1,\dots,K$ (here $G^r:\{1,\dots,N\} \mapsto \{1,\dots,K\}$ is the map between the stocks and the clusters),\footnote{\, For terminological definiteness here we focus on the level-1 clusters; it all straightforwardly applies to all levels. Also, the superscript $r$ in $\Omega^r_{ia}$ and $G^r(i)$ is an index, not a power.} be the binary loadings matrix from each run labeled by $r=1,\dots,M$. Here we are assuming that somehow we know how to properly order (i.e., align) the $K$ clusters from each run. This is a nontrivial assumption, which we will come back to momentarily. However, assuming, for a second, that we know how to do this, we can aggregate the loadings matrices $\Omega^r_{ia}$ into a single matrix ${\widetilde \Omega}_{ia} = \sum_{r=1}^M \Omega^r_{ia}$. Now, this matrix does not look like a binary loadings matrix. Instead, it is a matrix of occurrence counts, i.e., it counts how many times a given stock was assigned to a given cluster in the process of $M$ samplings. What we need to construct is a map $G$ such that one and only one stock belongs to each of the $K$ clusters. The simplest criterion is to map a given stock to the cluster in which ${\widetilde\Omega}_{ia}$ is maximal, i.e., where said stock occurs most frequently. A caveat is that there may be more than one such clusters. A simple criterion to resolve such an ambiguity is to assign said stock to the cluster with most cumulative occurrences (i.e., we take $q_a = \sum_{i=1}^N {\widetilde\Omega}_{ia}$ and assign this stock to the cluster with the largest $q_a$, if the aforesaid ambiguity occurs). In the unlikely event that there is still an ambiguity, we can try to do more complicated things, or we can simply assign such a stock to the cluster with the lowest value of the index $a$ -- typically, there is so much noise in the system that dwelling on such minutiae simply does not pay off.

{}However, we still need to tie up a loose end, to wit, our assumption that the clusters from different runs were somehow all aligned. In practice each run produces $K$ clusters, but i) they are not the same clusters and there is no foolproof way of mapping them, especially when we have a large number of runs; and ii) even if the clusters were the same or similar, they would not be ordered, i.e., the clusters from one run generally would be in a different order than clusters from another run.

{}So, we need a way to ``match" clusters from different samplings. Again, there is no magic bullet here either. We can do a lot of complicated and contrived things with not much to show for it at the end. A simple pragmatic solution is to use k-means to align the clusters from different runs. Each run labeled by $r=1,\dots,M$, among other things, produces a set of cluster centers $Y^r_{as}$. We can ``bootstrap" them by row into a $(KM) \times d$ matrix ${\widetilde Y}_{{\widetilde a}s} = Y^r_{as}$, where ${\widetilde a} = a + (r - 1)K$ takes values ${\widetilde a}=1,\dots,(KM)$. We can now cluster ${\widetilde Y}_{{\widetilde a}s}$ into $K$ clusters via k-means. This will map each value of ${\widetilde a}$ to $\{1,\dots,K\}$ thereby mapping the $K$ clusters from each of the $M$ runs to $\{1,\dots,K\}$. So, this way we can align all clusters. The ``catch" is that there is no guarantee that each of the $K$ clusters from each of the $M$ runs will be uniquely mapped to one value in $\{1,\dots,K\}$, i.e., we may have some empty clusters at the end of the day. However, this is fine, we can simply drop such empty clusters and aggregate (via the above procedure) the smaller number of $K^\prime < K$ clusters. I.e., at the end we will end up with an industry classification with $K^\prime$ clusters, which might be fewer than the target number of clusters $K$. This is not necessarily a bad thing. The dropped clusters might have been redundant in the first place. Another evident ``catch" is that even the number of resulting clusters $K^\prime$ is not deterministic. If we run this algorithm multiple times, we will get varying values of $K^\prime$. However, as we will see below, the aggregation procedure improves performance in our backtests and despite the variability in $K^\prime$ is also very stable from run to run. In Appendix \ref{app.A} we give the R source code for bottom-up clustering with various features we discuss above, including multilevel industry classification, the tweaks, and aggregation.\footnote{\, The source code in Appendix \ref{app.A}, Appendix \ref{app.B} and Appendix \ref{app.C} hereof is not written to be ``fancy" or optimized for speed or in any other way. Its sole purpose is to illustrate the algorithms described in the main text in a simple-to-understand fashion. See Appendix \ref{app.D} for some legalese.}

\subsubsection{Top-Down Clustering}\label{sub.topdown}

{}Above we discussed bottom-up clustering. We can go the other way around and do top-down clustering. I.e., we can construct a $P$-level classification as a sequence $K_P \rightarrow K_{P-1} \rightarrow \dots \rightarrow K_2 \rightarrow K_1$ (as before, $K_1 > K_2 > \dots > K_P$). More conveniently, we start with the entire universe of stocks and cluster ${\widehat R}_{is}$, $i=1,\dots,N$, into $L_P = K_P$ clusters. At level-$(P-1)$, we cluster each level-$P$ cluster $C_{a(P)} = \{i|i\in C_{a(P)}\}$, $a(P) = 1,\dots,K_P$, into $L_{P-1}$ clusters. We do this by clustering the returns ${\widehat R}_{is}$, $i\in C_{a(P)}$ via k-means into $L_{P-1}$ clusters.\footnote{\, More generally, we can nonuniformly cluster each level-$P$ cluster with its own $[L(a(P))]_{P-1}$.} At level-$(P-2)$, we cluster each level-$(P-1)$ cluster $C_{a(P-1)} = \{i|i\in C_{a(P-1)}\}$, $a(P-1) = 1,\dots,K_{P-1}$, into $L_{P-2}$ clusters. We do this by clustering the returns ${\widehat R}_{is}$, $i\in C_{a(P-1)}$ via k-means into $L_{P-2}$ clusters. And so on.\footnote{\, Note that, in contrast to bottom-up clustering, because here we are going ``backwards", it is convenient to  label the elements of each cluster at each level using the index $i$, which labels stocks.} In the zeroth approximation, $K_{P-1} = L_{P-1} K_P$, $K_{P-2} = L_{P-2} K_{P-1}$, and so on, so $K_1 = K_* = \prod_{\mu=1}^P L_\mu$. However, if at some level-$\mu$ we have some cluster $C_{a(\mu)}$ with $n_a(\mu) \leq L_\mu$, then we leave this cluster intact and do not cluster it, i.e., we ``roll it" forward unchanged. Therefore, we can have $K_1 < K_*$ at the most granular level-1. Also, instead of simply clustering via a single-sampling k-means, as above we can aggregate multiple samplings. Then at any level-$\mu$ we can end up clustering a given cluster $C_{a(\mu)}$ into $L_\mu$ or fewer clusters. Note, since here we work directly with the returns ${\widehat R}_{is}$, in contrast to the bottom-up approach, no cross-sectional demeaning is warranted at any level. In Appendix \ref{app.B} we give the R source code for top-down clustering, including with aggregation over multiple samplings.

\subsubsection{Relaxation Clustering}\label{relax}

{}Instead of k-means, which is nondeterministic, we can use other types of clustering, e.g., hierarchical agglomerative clustering. Let us focus on a 1-level classification here as we can always generalize it to multilevel cases as above. So, we have $N$ stocks, and we wish to cluster them into $K$ clusters. If $K$ is not preset, we can use SLINK \cite{SLINK}, etc. (see, e.g., \cite{HAC}). If we wish to preset $K$, then we can use a similar approach, except that it must be tweaked such that all observations are somehow squeezed into $K$ clusters. We give the R code for one such algorithm in Appendix \ref{app.C}. Basically, it is a relaxation algorithm which, as above, clusters ${\widehat R}_{is}$ (not $R_{is}$). The distance $D(i,j)$ between two $d$-vectors ${\widehat R}_{is}$ and ${\widehat R}_{js}$ is simply the Euclidean distance in ${\bf R}^d$. The initial cluster contains $i_1$ and $j_1$ with the smallest distance. If some $i_2$ and $j_2$ (such that $i_2\neq i_1$, $i_2\neq j_1$, $j_2\neq i_1$ and $j_2\neq j_1$) are such that $D(i_2,j_2)$ is smaller than the lesser of $D(i_1, \ell)$ and $D(j_1, \ell)$ for all $\ell$ ($\ell\neq i_1$ and $\ell\neq j_1$), then $i_2$ and $j_2$ form the second cluster. Otherwise $\ell$ that minimizes $D(i_1, \ell)$ or $D(j_1, \ell)$ is added to the first cluster. This is continued until there are $K$ clusters. Once we have $K$ clusters, we can only add to these clusters.\footnote{\, A brute force algorithm where at each step rows and columns are deleted from the matrix $D(i,j)$ is too slow. The R source code we give in Appendix \ref{app.C} is substantially more efficient than that. However, it is still substantially slower than the k-means based algorithms we discuss above.}

\section{Backtests}\label{sec.backtests}

{}Let us backtest the above algorithms for constructing statistical industry classification by utilizing the same backtesting procedure as in \cite{Het}. The remainder of this subsection very closely follows most parts of Section 6 thereof.\footnote{\, We ``rehash" it here not to be repetitive but so that our presentation here is self-contained.}

\subsection{Notations}

{}Let $P_{is}$ be the time series of stock prices, where $i=1,\dots,N$ labels the stocks, and $s=1,2,\dots$ labels the trading dates, with $s=1$ corresponding to the most recent date in the time series. The superscripts $O$ and $C$ (unadjusted open and close prices) and $AO$ and $AC$ (open and close prices fully adjusted for splits and dividends) will distinguish the corresponding prices, so, e.g., $P^C_{is}$ is the unadjusted close price. $V_{is}$ is the unadjusted daily volume (in shares). Also, for each date $s$ we define the overnight return as the previous-close-to-open return:
\begin{equation}\label{c2o.ret}
 E_{is} = \ln\left({P^{AO}_{is} / P^{AC}_{i,s+1}}\right)
\end{equation}
This return will be used in the definition of the expected return in our mean-reversion alpha. We will also need the close-to-close return
\begin{equation}\label{c2c.ret}
 R_{is} = \ln\left({P^{AC}_{is} / P^{AC}_{i,s+1}}\right)
\end{equation}
An out-of-sample (see below) time series of these returns will be used in constructing the risk models. All prices in the definitions of $E_{is}$ and $R_{is}$ are fully adjusted.

{}We assume that: i) the portfolio is established at the open\footnote{\, This is a so-called ``delay-0" alpha: the same price, $P^O_{is}$ (or adjusted $P^{AO}_{is}$), is used in computing the expected return (via $E_{is}$) and as the establishing fill price.} with fills at the open prices $P^O_{is}$; ii) it is liquidated at the close on the same day -- so this is a purely intraday alpha -- with fills at the close prices $P^C_{is}$; and iii) there are no transaction costs or slippage -- our aim here is not to build a realistic trading strategy, but to test {\rm relative} performance of various statistical industry classifications. The P\&L for each stock
\begin{equation}
 \Pi_{is} = H_{is}\left[{P^C_{is}\over P^O_{is}}-1\right]
\end{equation}
where $H_{is}$ are the {\em dollar} holdings. The shares bought plus sold (establishing plus liquidating trades) for each stock on each day are computed via $Q_{is} = 2 |H_{is}| / P^O_{is}$.

\subsection{Universe Selection}\label{sub.univ}

{}For the sake of simplicity,\footnote{\, In practical applications, the trading universe of liquid stocks typically is selected based on market cap, liquidity (ADDV), price and other (proprietary) criteria.} we select our universe based on the average daily dollar volume (ADDV) defined via (note that $A_{is}$ is out-of-sample for each date $s$):
\begin{equation}\label{ADDV}
 A_{is}= {1\over m} \sum_{r=1}^m V_{i, s+r}~P^C_{i, s+r}
\end{equation}
We take $m=21$ (i.e., one month), and then take our universe to be the top 2000 tickers by ADDV. To ensure that we do not inadvertently introduce a universe selection bias, we rebalance monthly (every 21 trading days, to be precise). I.e., we break our 5-year backtest period (see below) into 21-day intervals, we compute the universe using ADDV (which, in turn, is computed based on the 21-day period immediately preceding such interval), and use this universe during the entire such interval. We do have the survivorship bias as we take the data for the universe of tickers as of 9/6/2014 that have historical pricing data on http://finance.yahoo.com (accessed on 9/6/2014) for the period 8/1/2008 through 9/5/2014. We restrict this universe to include only U.S. listed common stocks and class shares (no OTCs, preferred shares, etc.) with BICS (Bloomberg Industry Classification System) sector assignments as of 9/6/2014.\footnote{\, The choice of the backtesting window is intentionally taken to be exactly the same as in \cite{Het} to simplify various comparisons, which include the results therefrom.} However, as discussed in detail in Section 7 of \cite{MeanRev}, the survivorship bias is not a leading effect in such backtests.\footnote{\, Here we are after the {\em relative outperformance}, and it is reasonable to assume that, to the leading order, individual performances are affected by the survivorship bias approximately equally as the construction of all alphas and risk models is ``statistical" and oblivious to the universe.}

\subsection{Backtesting}\label{sub.back}

{}We run our simulations over a period of 5 years (more precisely, 1260 trading days going back from 9/5/2014, inclusive). The annualized return-on-capital (ROC) is computed as the average daily P\&L divided by the intraday investment level $I$ (with no leverage) and multiplied by 252. The annualized Sharpe Ratio (SR) is computed as the daily Sharpe ratio multiplied by $\sqrt{252}$. Cents-per-share (CPS) is computed as the total P\&L in cents (not dollars) divided by the total shares traded.

\subsection{Optimized Alphas}\label{sub.opt}

{}The optimized alphas are based on the expected returns $E_{is}$ optimized via Sharpe ratio maximization using heterotic risk models \cite{Het} based on statistical industry classifications we are testing.\footnote{\, In \cite{Het} BICS is used for the industry classification. Here we simply plug in the statistical industry classification instead of BICS. In the case of a single-level industry classification, we can either add the second level consisting of the ``market" with the $N\times 1$ unit matrix as the loadings matrix; or, equivalently, we can use the option {\tt{\small{mkt.fac = T}}} in the R function {\tt{\small{qrm.het()}}} in Appendix B of \cite{Het}, which accomplishes this internally.} We compute the heterotic risk model covariance matrix $\Gamma_{ij}$ every 21 trading days (same as for the universe). For each date (we omit the index $s$) we maximize the Sharpe ratio subject to the dollar neutrality constraint:
\begin{eqnarray}
 &&{\cal S} = {\sum_{i=1}^N H_i~E_i\over {\sqrt{\sum_{i,j=1}^N {\Gamma}_{ij}~H_i~H_j}}} \rightarrow \mbox{max}\\
 &&\sum_{i=1}^N H_i = 0\label{d.n.opt}
\end{eqnarray}
In the absence of bounds, the solution is given by
\begin{equation}\label{H.opt}
 H_i = -\eta \left[\sum_{j = 1}^N {\Gamma}^{-1}_{ij}~E_j - \sum_{j=1}^N {\Gamma}^{-1}_{ij}~{{\sum_{k,l=1}^N {\Gamma}^{-1}_{kl}~E_l}\over{\sum_{k,l = 1}^N {\Gamma}^{-1}_{kl}}}\right]
\end{equation}
where ${\Gamma}^{-1}$ is the inverse of ${\Gamma}$, and $\eta > 0$ (mean-reversion alpha) is fixed via (we set the investment level $I$ to \$20M in our backtests)
\begin{equation}
 \sum_{i=1}^N \left|H_i\right| = I
\end{equation}
Note that (\ref{H.opt}) satisfies the dollar neutrality constraint (\ref{d.n.opt}).

{}In our backtests we impose position bounds (which in this case are the same as trading bounds as the strategy is purely intraday) in the Sharpe ratio maximization:
\begin{equation}\label{liq}
 |H_{is}| \leq 0.01~A_{is}
\end{equation}
where $A_{is}$ is ADDV defined in (\ref{ADDV}). In the presence of bounds computing $H_i$ requires an iterative procedure and we use the R code in Appendix C of \cite{Het}.

\subsection{Simulation Results}

{}Table \ref{table.try1.100.30.10} summarizes simulation results for 11 independent runs for the ``bottom-up" 3-level statistical industry classification with $K_1 = 100$, $K_2 = 30$ and $K_3 = 10$ (see Subsection \ref{sub.bottomup}). Despite the nondeterministic nature of the underlying k-means algorithm, pleasantly, the backtest results are very stable. Table \ref{table.try100.100} summarizes simulation results for 11 independent runs for the ``bottom-up" single-level statistical industry classification with the target number of clusters $K=100$ based on aggregating 100 samplings (so the actual number of resultant clusters $K^\prime$ can be smaller than $K$ -- see Subsection \ref{sub.aggr}). Again, the backtest results are very stable. Table \ref{table.try100.100.30.10} summarizes simulation results for 23 independent runs for the ``bottom-up" 3-level statistical industry classification with the target number of clusters $K_1=100$, $K_2 = 30$ and $K_3 = 10$ based on aggregating 100 samplings (so the actual number of resultant clusters $K_\mu^\prime$ can be smaller than $K_\mu$, $\mu=1,2,3$ -- see Subsection \ref{sub.aggr}). The first 15 (out of 23) runs correspond to {\tt{\small{norm.cl.ret = F}}} (this corresponds to choice (i) after Equation (\ref{Rmu}) in Subsection \ref{sub.bottomup}), while the other 8 runs correspond to {\tt{\small{norm.cl.ret = T}}} (this corresponds to choice (ii) after said Equation); see the function {\tt{\small{qrm.stat.ind.class.all()}}} in Appendix \ref{app.A}. The aforesaid stability persists to these cases as well. Table \ref{table.num.clust} summarizes the number of actual clusters in a statistical industry classification obtained via aggregating 100 samplings. The target numbers of clusters in a 3-level hierarchy are $K_1 = 100$, $K_2 = 30$ and $K_3 = 10$, as in Table \ref{table.try100.100.30.10}.

{}Table \ref{table.try1.topdown} summarizes simulation results for ``top-down" 3-level statistical industry classifications obtained via a single sampling in each run, with 3 runs for each $L_\mu$. The 3-vector $L_\mu$, $\mu=1,2,3$, is defined in Subsection \ref{sub.topdown}. Recall that in the zeroth approximation the number of clusters at the most granular level-1 is $K_1 = L_1 L_2 L_3$; however, the actual value can be lower due to the reasons explained in Subsection \ref{sub.topdown}. Here too we observe substantial stability. Table \ref{table.try100.topdown} summarizes simulation results for ``top-down" 3-level statistical industry classifications obtained via aggregating 100 samplings in each run, with 3 runs for each $L_\mu$. Stability persists.

{}From the above results it is evident that aggregating multiple samplings on average improves both performance and stability. Furthermore, not surprisingly, decreasing granularity worsens the Sharpe ratio. 3-level classifications outperform single-level classifications.\footnote{\, Also, ``bottom-up" by construction uses more information than and outperforms ``top-down".} Above we mentioned that clustering ${\widehat R}_{is} = R_{is}/\sigma_i^2$ outperforms clustering ${\widetilde R}_{is} = R_{is}/\sigma_i$, which in turn outperforms clustering $R_{is}$. Thus, a random run for the ``bottom-up" 3-level classification with $K_1 = 100$, $K_2 = 30$ and $K_3 = 10$ based on clustering $R_{is}$ using a single sampling produced a typical performance with ROC = 41.885\%, SR = 15.265 and CPS = 1.889 (cf. Table \ref{table.try1.100.30.10}). A random run for the ``bottom-up" 3-level classification with $K_1 = 100$, $K_2 = 30$ and $K_3 = 10$ based on clustering ${\widetilde R}_{is}$ using a single sampling produced a typical performance with ROC = 42.072\%, SR = 15.840 and CPS = 1.973 (cf. Table \ref{table.try1.100.30.10}).\footnote{\, Table \ref{table.try1.100.30.10} is based on clustering ${\widehat R}_{is}$ defined via (\ref{tweak}). However, clustering ${\widehat R}^*_{is} = R_{is}/\sigma^2_i$ produces essentially the same results. Thus, a random run for the ``bottom-up" 3-level classification with $K_1 = 100$, $K_2 = 30$ and $K_3 = 10$ based on clustering ${\widehat R}^*_{is}$ via aggregating 100 samplings produced a typical performance with ROC = 41.707\%, SR = 16.220 and CPS = 2.091 (cf. Table \ref{table.try100.100.30.10}).}

{}In contrast to nondeterministic k-means based algorithms, the relaxation algorithm (Subsection \ref{relax}) is completely deterministic. We run it using the code in Appendix \ref{app.C} to obtain a 3-level classification with the target numbers of clusters $K_1 = 100$, $K_2 = 30$ and $K_3 = 10$ (as in the ``bottom-up" cases, we cross-sectionally demean the level-2 and level-3 returns, but not the level-1 returns). The simulation results are sizably worse than for k-means based algorithms: ROC = 41.266\%, SR = 15.974 and CPS = 1.990. How come? Intuitively, this is not surprising. All such relaxation mechanisms (hierarchical agglomerative algorithms) start with a ``seed", i.e., the initial cluster picked based on some criterion. In Subsection \ref{relax} this is the first cluster containing the pair $(i_1,j_1)$ that minimized the Euclidean distance. However, generally this choice is highly unstable out-of-sample, hence underperformance. In contrast, k-means is much more ``statistical", especially with aggregation.

\section{How to Fix Cluster Numbers?}\label{sec.dyn}

{}Thus far we have picked the number of clusters $K_\mu$ as well as the number of levels $P$ ``ad hoc".\footnote{\, Here we focus on the k-means based ``bottom-up" and ``top-down" algorithms. As discussed above, the relaxation algorithm underperforms the k-means based algorithms.} Can we fix them ``dynamically"? If we so choose, here we can do a lot of complicated things. Instead, our approach will be based on pragmatism (rooted in financial considerations) and simplicity.\footnote{\, A variety of methods for fixing the number of clusters have been discussed in other contexts. See, e.g., \cite{Rousseeuw}, \cite{Goutte}, \cite{Sugar}, \cite{Lleiti}, \cite{DeAmorim}.} As can be surmised from Tables \ref{table.try100.100} and \ref{table.try100.100.30.10}, the number of levels does not make it or break it in our context. What is more important is the number of clusters. So, suppose we have a given number of levels $P > 1$. Let us start by asking, what should $K_1$ (most granular level) and $K_P$ (least granular level) be? In practice, the number of stocks $N>d-1$, so the sample correlation matrix $\Psi_{ij}$ is singular. (In fact, in most practical applications $N\gg d-1$.) We can model it via statistical risk models \cite{StatRM}. These are factor models obtained by truncating the spectral decomposition of $\Psi_{ij}$
\begin{equation}
 \Psi_{ij} = \sum_{a = 1}^{d - 1} \lambda^{(a)}~V_i^{(a)}~V_j^{(a)}
\end{equation}
via the first $d - 1$ principal components $V_i^{(a)}$ (only $d - 1$ eigenvalues $\lambda^{(a)}$ are positive, $\lambda^{(1)} > \lambda^{(2)} > \dots, \lambda^{(d-1)} > 0$, while the rest of the eigenvalues $\lambda^{(a)} \equiv 0$, $a\geq d$) to the first $F$ principal components ($F < d - 1$) and compensating the deficit on the diagonal (as $\Psi_{ii}\equiv 1$) by adding diagonal specific (idiosyncratic) variance $\xi_i^2$:
\begin{equation}
 \Gamma_{ij} = \xi_i^2~\delta_{ij} + \sum_{a = 1}^F \lambda^{(a)}~V_i^{(a)}~V_j^{(a)}
\end{equation}
I.e., we approximate $\Psi_{ij}$ (which is singular) via $\Gamma_{ij}$ (which is positive-definite as all $\xi_i^2 > 0$ and are fixed from the requirement that $\Gamma_{ii} \equiv 1$). The question then is, what should $F$ be? One simple (``minimization" based) algorithm for fixing $F$ is given in \cite{Het}. Another, even simpler algorithm recently proposed in \cite{StatRM}, is based on eRank (effective rank) defined below.\footnote{\, For prior works on fixing $F$, see, e.g., \cite{Connor} and \cite{Bai}.}

\subsection{Effective Rank}

{}Thus, we simply set (here $\mbox{Round}(\cdot)$ can be replaced by $\mbox{floor}(\cdot) = \lfloor\cdot\rfloor$)
\begin{equation}\label{eq.eRank}
 F = \mbox{Round}(\mbox{eRank}(\Psi))
\end{equation}
Here $\mbox{eRank}(Z)$ is the effective rank \cite{RV} of a symmetric semi-positive-definite (which suffices for our purposes here) matrix $Z$. It is defined as
\begin{eqnarray}
 &&\mbox{eRank}(Z) = \exp(H)\\
 &&H = -\sum_{a=1}^L p_a~\ln(p_a)\\
 &&p_a = {\lambda^{(a)} \over \sum_{b=1}^L \lambda^{(b)}}
\end{eqnarray}
where $\lambda^{(a)}$ are the $L$ {\em positive} eigenvalues of $Z$, and $H$ has the meaning of the (Shannon a.k.a. spectral) entropy \cite{Campbell60}, \cite{YGH}.

{}The meaning of $\mbox{eRank}(Z)$ is that it is a measure of the effective dimensionality of the matrix $Z$, which is not necessarily the same as the number $L$ of its positive eigenvalues, but often is lower. This is due to the fact that many returns can be highly correlated (which manifests itself by a large gap in the eigenvalues) thereby further reducing the effective dimensionality of the correlation matrix.

\subsection{Fixing $K_\mu$}\label{sub.dyn}

{}There is no magic bullet here. It just has to make sense. Intuitively, it is natural to identify the number of clusters $K_P$ at the least granular level with the number of factors $F$ in the context of statistical risk models.\footnote{\, The number of factors $F$ essentially measures the effective number of degrees of freedom in the underlying time series of returns $R_{is}$. Hence identification of $K_P$ with this number.} In the following, we will therefore simply take
\begin{equation}
 K_P = \mbox{Round}(\mbox{eRank}(\Psi))
\end{equation}
Adding more granular levels explores deeper substructures in the time series of returns based on the closeness criterion. In this regard, we can fix the number of clusters $K_1$ at the most granular level as follows. The average number of stocks per cluster at level-1 is $N_1 = N/K_1$ (we are being cavalier with rounding). Assume for a second that the number of stocks in each cluster is the same and equal $N_1$. If $N_1 > d-1$, then the sub-matrices $\Psi_{ij}$, $i,j\in C_{a(1)}$ (recall that $C_{a(1)}$, $a(1) = 1,\dots,K_1$, are the level-1 clusters) are singular. For $N_1 \leq d - 1$ they are nonsingular. Therefore, intuitively, it is natural to fix $K_1$ by requiring that $N_1 = d - 1$. Restoring rounding, in the following we will set
\begin{equation}
 K_1 = \mbox{Round}(N / (d-1))
\end{equation}
What about $K_\mu$, $1 < \mu < P$? Doing anything overly complicated here would be overkill. Here is a simple prescription (assuming $K_1 > K_P$):\footnote{\, I.e., $K_\mu$ are (up to rounding) equidistant on the log scale. For $P=3$ the ``midpoint" $K_2 = \sqrt{K_1 K_P}$ is simply the geometric mean. With this prescription, we can further fix $P$ via some heuristic, e.g., take maximal $P$ such that the difference $K_{P-1} - K_P\geq \Delta$, where $\Delta$ is preset, say, $\Delta = K_P$. For $K_1 = 100$ and $K_P = 10$, this would give us $P=4$ with $K_2 = 46$ and $K_3 = 22$.}
\begin{equation}
 K_\mu = \left[K_1^{P - \mu}~K_P^{\mu - 1}\right]^{1\over{P-1}},~~~\mu=1,\dots,P
\end{equation}
We give the R source code for building ``bottom-up" statistical industry classifications using this prescription in Appendix \ref{app.A}. Table \ref{table.dyn} summarized simulation results for $P =2,3,4,5$. It is evident that the number of levels is not a driver here. The results are essentially the same as for $K_1 = 100$ (recall that $N = 2000$ and $d = 21$ in our case) in Tables \ref{table.try100.100} and \ref{table.try100.100.30.10}. Table \ref{table.try100.K} isolates the $K$ dependence and suggests that the performance peaks around $K = 100$. Again, there is no magic bullet here.\footnote{\, Note from Table \ref{table.try100.K} that too little granularity lowers the Sharpe ratio due to insufficient coverage of the risk space, while too much granularity lowers cents-per-share due to overtrading.}

\subsection{Comparisons}

{}Let us compare the (very stable) results we obtained for statistical industry classifications with two ``benchmarks": statistical risk models \cite{StatRM} and heterotic risk models with BICS used as the industry classification \cite{Het}. More precisely, statistical risk models in \cite{StatRM} were built based on the sample correlation matrix $\Psi_{ij}$, which is equivalent to basing them on normalized returns ${\widetilde R}_{is} = R_{is}/\sigma_i$. If we use the eRank based algorithm for fixing the number of statistical risk factors $F$, then the performance is ROC = 40.777\%, SR = 14.015 and CPS = 1.957 \cite{StatRM}.\footnote{\, In \cite{StatRM} rounding is to 2 decimals, while here we round to 3 decimals.} However, as we argued above, it makes more sense to build models using ${\widehat R}_{is} = R_{is}/\sigma_i^2$. So, we should compare our results here with the statistical risk models based on ${\widehat R}_{is}$. To achieve this, we can simply replace the line {\tt{\small{tr <- apply(ret, 1, sd)}}} in the R function {\tt{\small{qrm.erank.pc(ret, use.cor = T)}}} in Appendix A of \cite{StatRM} by {\tt{\small{tr <- apply(ret, 1, sd) / apply(qrm.calc.norm.ret(ret), 1, sd)}}}, where the R function {\tt{\small{qrm.calc.norm.ret()}}} is given in Appendix \ref{app.A} hereof. The performance is indeed better: ROC = 40.878\%, SR = 14.437 and CPS = 2.018. So, the k-means based clustering algorithms still outperform statistical risk models, which implies that going beyond the $F$ statistical factors adds value, i.e., there is more structure in the data than is captured by the principal components alone. However, statistical industry classifications still sizably underperform heterotic risk models based on BICS \cite{Het}:\footnote{\, Here we use the results from \cite{HetPlus}, which slightly differ from those in \cite{Het}, where rounding down (as opposed to simply rounding) was employed.} ROC = 49.005\%, SR = 19.230 and CPS = 2.365. Clearly, statistical industry classifications are not quite on par with industry classifications such as BICS, which are based on fundamental/economic data (such as companies' products and services and more generally their revenue sources, suppliers, competitors, partners, etc.). Such industry classifications are essentially independent of the pricing data and, if well-built, tend to be rather stable out-of-sample as companies seldom jump industries. In contrast, statistical industry classifications by nature are less stable out-of-sample. However, they can add substantial value when ``fundamental" industry classifications are unavailable, including for returns other than for stocks, e.g., quantitative trading alphas \cite{Billion}.

{}Finally, before we close this section, let us discuss the ``top-down" classifications with dynamically determined numbers of clusters $K_\mu$. More precisely, recall that in this case we work with the vector $L_\mu$ (see Subsection \ref{sub.topdown}). The code we used in the ``bottom-up" case (Appendix \ref{app.A}) can be used in this case as well (via a parameter choice). A random (and typical) run with $P=3$ gives ROC = 41.657\%, SR = 15.897 and CPS = 2.079, while another such run with $P=4$ gives ROC = 41.683\%, SR = 15.697 and 2.073. These results are in line with our results in Table \ref{table.try100.topdown}.

\section{Hybrid Industry Classification}\label{sec.hybrid}

{}One application of a statistical industry classification is to use it as a means for improving a ``fundamental" industry classification such as BICS, GICS, etc. Thus, a ``fundamental" classification at the most granular level can have overly large sub-industries, using the BICS nomenclature for definiteness. One way to deal with such large sub-industries is to further cluster them using statistical industry classification methods discussed above. Let us illustrate this using BICS as an example.

{}Table \ref{table.bics.summary} summarizes top 10 most populous (by stock counts) sub-industries in one of our 2000 stock backtesting portfolios. For comparison, the stock count summary across all 165 sub-industries in this sample is Min = 1, 1st Qu. = 3, Median = 8, Mean = 12.12, 3rd Qu. = 15, Max = 94, StDev = 14.755, MAD = 8.896 (see Table \ref{table.num.clust} for notations). So, we have some ``large" sub-industries, which are outliers.

{}We can further split these large sub-industries into smaller clusters using our ``bottom-up" clustering algorithm. In fact, it suffices to split them using a single-level algorithm. We give the R code for improving an existing ``fundamental" industry classification using our statistical industry classification algorithm in Appendix \ref{app.A}. The idea is simple. Let us label the sub-industries (the most granular level) in the ``fundamental" industry classification via $A = 1,\dots, K_*$. Let $N_A$ be the corresponding stock counts. Let
\begin{equation}
 \kappa_A = \mbox{Round}(N_A / (d-1))
\end{equation}
We then split each sub-industry with $\kappa_A > 1$ into $\kappa_A$ clusters. Table \ref{table.improve} summarizes the simulation results for 14 runs. This evidently improves performance. Table \ref{table.ind.counts} gives summaries of top 10 most populous sub-industries before and after statistical clustering based on 60 datapoints at the end of each 21-trading-day interval in our backtests (recall that we have $1260 = 60 \times 21$ trading days -- see Section \ref{sec.backtests}). The average numbers of sub-industries are 165.45 before and 184.1 after clustering.

\section{Concluding Remarks}\label{sec.concl}

{}In this paper we discuss all sorts of nuances in constructing statistical industry classifications. Under the hood we have clustering algorithms. However, it is precisely those nuances that make a sizable difference. E.g., if we naively cluster $R_{is}$, we get a highly suboptimal result compared with clustering ${\widetilde R}_{is} = R_{is}/\sigma_i$, which in turn underperforms clustering ${\widehat R}_{is} = R_{is}/\sigma_i^2$. In this regard, let us tie up a ``loose end" here: what if we cluster ${\overline R}_{is} = R_{is}/\sigma_i^3$? It underperforms clustering ${\widehat R}_{is}$. Thus, a typical run for a 3-level ``bottom-up" classification with target cluster numbers $K_1 = 100$, $K_2 = 30$ and $K_3 = 10$ based on clustering ${\overline R}_{is}$ and aggregating 100 samplings produces the following: ROC = 40.686, SR = 15.789 and CPS = 2.075.

{}So, suppressing returns $R_{is}$ by $\sigma_i^2$ indeed appears to be optimal -- for the intuitive reasons we discussed above. We saw the same in the context of statistical risk models. In this regard, it would be interesting to explore this avenue in the context of heterotic risk models \cite{Het} and the more general (heterotic CAPM) construction of \cite{HetPlus}. In the latter framework, it would be interesting to utilize an aggregated (non-binary) matrix ${\widetilde \Omega}_{ia}$ (see Subsection \ref{sub.aggr}). These ideas are outside of the scope hereof and we hope to return to them elsewhere.

\appendix
\section{R Code for Bottom-Up Clustering}\label{app.A}
\subsection{Code for Single-Level Clustering}\label{app.A.main}

{}In this subsection we give the R source code (R Package for Statistical Computing, http://www.r-project.org) for single-level ``bottom-up" clustering (see Subsection \ref{sub.bottomup}). The code is straightforward and self-explanatory as it simply follows the formulas and logic in Section \ref{sec.stat.clust}. The main function is {\tt{\small{qrm.stat.ind.class(ret, k, iter.max = 10, num.try = 100, demean.ret = F)}}}, which internally calls two auxiliary functions. The function {\tt{\small{qrm.calc.norm.ret(ret)}}} normalizes the $N\times d$ matrix {\tt{\small{ret}}} (the return time series $R_{is}$, $i = 1,\dots,N$, $s=1,\dots,d$) following Subsection \ref{sub.norm.ret} and outputs ${\widehat R}_{is}$ (see Eq. (\ref{tweak})). The inputs of the function {\tt{\small{qrm.calc.kmeans.ind(x, centers, iter.max)}}} are the same as in the built-in R function {\tt{\small{kmeans()}}}, and it outputs a list: {\tt{\small{res\$ind}}} is the $N\times K$ binary industry classification matrix $\Omega_{ia} = \delta_{G(i), a}$, where $G:\{1,\dots,N\}\mapsto\{1,\dots,K\}$ maps stocks to $K$ clusters labeled by $a = 1,\dots,K$; {\tt{\small{res\$centers}}} is the $K \times d$ matrix $Y_{as}$ of the cluster centers; {\tt{\small{res\$cluster}}} is an $N$-vector $G(i)$; and the number of clusters $K$ is passed into this function via the argument {\tt{\small{centers}}} as in {\tt{\small{kmeans()}}}. The inputs of the function {\tt{\small{qrm.stat.ind.class()}}} are: {\tt{\small{ret}}} defined above; the target number of clusters {\tt{\small{k}}}; the maximum number of k-means iterations {\tt{\small{iter.max}}} (same as in {\tt{\small{kmeans()}}}) with the default {\tt{\small{iter.max = 10}}}, however, in all our backtests we set {\tt{\small{iter.max = 100}}} (with 100\% convergence rate); {\tt{\small{num.try = 100}}} (default), which is the number of independent k-means samplings to be aggregated (see Subsection \ref{sub.aggr}), with {\tt{\small{num.try = 1}}} corresponding to no aggregation (i.e., a single k-means sampling); {\tt{\small{demean.ret = F}}} (default) corresponds to taking vanilla $R_{is}$, while {\tt{\small{demean.ret = T}}} corresponds to demeaning it cross-sectionally before running the rest of the code (see Subsection \ref{sub.demean}). The main function outputs the $N\times K^\prime$ binary industry classification matrix ($K^\prime\leq K$).\footnote{\, Recall from Subsection \ref{sub.aggr} that $K^\prime$ can be less than $K$ unless {\tt{\small{num.try = 1}}}.}\\
\\
{\tt{\small
\noindent qrm.calc.norm.ret <- function (ret)\\
\{\\
\indent s <- apply(ret, 1, sd)\\
\indent u <- log(s)\\
\indent u <- u - (median(u) - 3 * mad(u))\\
\indent u <- exp(u)\\
\indent take <- u > 1\\
\indent u[!take] <- 1\\
\indent x <- ret / s / u\\
\indent return(x)\\
\}\\
\\
\noindent qrm.calc.kmeans.ind <- function (x, centers, iter.max)\\
\{\\
\indent res <- new.env()\\
\indent y <- kmeans(x, centers, iter.max = iter.max)\\
\indent x <- y\$cluster\\
\indent k <- nrow(y\$centers)\\
\indent z <- matrix(NA, length(x), k)\\
\indent for(j in 1:k)\\
\indent \indent z[, j] <- as.numeric(x == j)\\
\indent z <- z[, colSums(z) > 0]\\
\indent res\$ind <- z\\
\indent res\$centers <- y\$centers\\
\indent res\$cluster <- y\$cluster\\
\indent return(res)\\
\}\\
\\
\noindent qrm.stat.ind.class <- function (ret, k,\\
\indent iter.max = 10, num.try = 100, demean.ret = F)\\
\{\\
\indent if(demean.ret)\\
\indent \indent ret <- t(t(ret) - colMeans(ret))\\
\\
\indent norm.ret <- qrm.calc.norm.ret(ret)\\
\\
\indent for(i in 1:num.try)\\
\indent \{\\
\indent \indent res <- qrm.calc.kmeans.ind(norm.ret, k, iter.max)\\
\\
\indent \indent if(num.try == 1)\\
\indent \indent \indent return(res\$ind)\\
\\
\indent \indent if(i == 1)\\
\indent \indent \{\\
\indent \indent \indent comb.cent <- res\$centers\\
\indent \indent \indent comb.ind <- res\$ind\\
\indent \indent \}\\
\indent \indent else\\
\indent \indent \{\\
\indent \indent \indent comb.cent <- rbind(comb.cent, res\$centers)\\
\indent \indent \indent comb.ind <- cbind(comb.ind, res\$ind)\\
\indent \indent \}\\
\indent \}\\
\\
\indent res <- qrm.calc.kmeans.ind(comb.cent, k, iter.max)\\
\indent cl <- res\$cluster\\
\indent z <- matrix(0, nrow(ret), k)\\
\\
\indent for(i in 1:length(cl))\\
\indent \indent z[, cl[i]] <- z[, cl[i]] + comb.ind[, i]\\
\\
\indent q <- colSums(z)\\
\indent for(i in 1:nrow(z))\\
\indent \{\\
\indent \indent take <- z[i, ] == max(z[i, ])\\
\indent \indent take <- take \& q == max(q[take])\\
\indent \indent ix <- 1:ncol(z)\\
\indent \indent ix <- min(ix[take])\\
\indent \indent z[i, ] <- 0\\
\indent \indent z[i, ix] <- 1\\
\indent \}\\
\indent z <- z[, colSums(z) > 0]\\
\indent return(z)\\
\}
}}

\subsection{Code for Multilevel Clustering}\label{app.A.all}

{}In this subsection we give the R source code for building multilevel ``bottom-up" statistical industry classifications (see Subsection \ref{sub.bottomup}). There is only one function {\tt{\small{qrm.stat.ind.class.all(ret, k, iter.max = 10, num.try = 100, do.demean = rep(F, length(k)), norm.cl.ret = F)}}}, which internally calls the main function {\tt{\small{qrm.stat.ind.class()}}} from Subsection \ref{app.A.main} with the same inputs {\tt{\small{ret}}}, {\tt{\small{iter.max}}} and {\tt{\small{num.try}}}, and the following new inputs: {\tt{\small{k}}} is a $P$-vector $K_\mu$, $\mu=1,\dots,P$, where $P$ is the number of levels (see Subsection \ref{sub.bottomup}); {\tt{\small{do.demean = rep(F, length(k))}}} is a Boolean $P$-vector, which sets the input {\tt{\small{demean.ret}}} in {\tt{\small{qrm.stat.ind.class()}}} (in our backtests we set all elements of {\tt{\small{do.demean}}} to {\tt{\small{TRUE}}} except for the first one); {\tt{\small{norm.cl.ret = F}}} corresponds to choice (i) right after Eq. (\ref{Rmu}), and {\tt{\small{norm.cl.ret = T}}} corresponds to choice (ii) (we mostly use choice (i) in our backtest -- see Section \ref{sec.backtests}). The output is a list: {\tt{\small{ind.list[[i]]}}} is the $N \times K_\mu$ ({\em not} $K_{{\mu}-1}\times K_\mu$) binary industry classification matrix at level {\tt{\small{i}}} = $\mu$, i.e., it maps stocks to the level-$\mu$ clusters $C_{a(\mu)}$.\\
\\
{{\tt{\small
\noindent qrm.stat.ind.class.all <- function (ret, k,\\
\indent iter.max = 10, num.try = 100,\\
\indent \indent do.demean = rep(F, length(k)), norm.cl.ret = F)\\
\{\\
\indent ind.list <- list()\\
\\
\indent for(i in 1:length(k))\\
\indent \{\\
\indent \indent ind.list[[i]] <- qrm.stat.ind.class(ret, k[i],\\
\indent \indent \indent iter.max = iter.max, num.try = num.try,\\
\indent \indent \indent \indent demean.ret = do.demean[i])\\
\indent \indent if(norm.cl.ret)\\
\indent \indent \indent ret <- t(ind.list[[i]]) \%*\% qrm.calc.norm.ret(ret)\\
\indent \indent else\\
\indent \indent \indent ret <- t(ind.list[[i]]) \%*\% ret\\
\\
\indent \indent if(i > 1)\\
\indent \indent \{\\
\indent \indent \indent ind.list[[i]] <- ind.list[[i - 1]] \%*\% ind.list[[i]]\\
\indent \indent \indent take <- ind.list[[i]] > 0\\
\indent \indent \indent ind.list[[i]][take] <- 1\\
\indent \indent \}\\
\indent \}\\
\indent return(ind.list)\\
\}
}}
\subsection{Code for Dynamically Fixing Cluster Numbers}

{}In this subsection we give the R source code for building multilevel ``bottom-up" statistical industry classifications with the numbers of clusters fixed dynamically (see Section \ref{sec.dyn} and Subsection \ref{sub.dyn}). The main function {\tt{\small{qrm.stat.ind.class.dyn(ret, p, iter.max = 10, num.try = 100, top.down = F)}}} has the same inputs as above except: {\tt{\small{p}}} is the number of levels, and when {\tt{\small{top.down = F}}} it internally calls the function {\tt{\small{qrm.stat.ind.class.all()}}} from Subsection \ref{app.A.all}, while when {\tt{\small{top.down = T}}} it internally calls the function {\tt{\small{qrm.stat.class()}}} from Appendix \ref{app.B}. The main function internally also calls the function {\tt{\small{qrm.eigen(ret, calc.cor = T)}}}, which provides a more efficient way of computing eigenpairs of the sample covariance (when {\tt{\small{calc.cor = F}}}) or correlation (when {\tt{\small{calc.cor = T}}}) matrix based on {\tt{\small{ret}}} than the built-in R function {\tt{\small{eigen()}}} by internally calling the R function {\tt{\small{qrm.calc.eigen.eff(ret, calc.cor = F)}}} from Appendix C of \cite{StatRM} (when $d \leq N+1$). It also internally calls the within R function {\tt{\small{qrm.calc.cov.mat(x, calc.cor = F)}}} (when $d > N+1$). The output is a list {\tt{\small{ind.list}}}, same as in Subsection \ref{app.A.all}.\\
\\
{{\tt\small
\noindent qrm.stat.ind.class.dyn <- function (ret, p,\\
\indent iter.max = 10, num.try = 100, top.down = F)\\
\{\\
\indent k1 <- round(nrow(ret) / (ncol(ret) - 1))\\
\indent if(p > 1)\\
\indent \{\\
\indent \indent y <- qrm.eigen(ret, calc.cor = T)\$values\\
\indent \indent kp <- round(qrm.calc.erank(y, F))\\
\\
\indent \indent if(k1 < kp)\\
\indent \indent \indent p <- 1\\
\indent \}\\
\\
\indent if(p == 1)\\
\indent \indent k <- k1\\
\indent else\\
\indent \{\\
\indent \indent q <- ((p - 1):0) / (p - 1)\\
\indent \indent k <- round(k1\^{}q * kp\^{}(1 - q))\\
\indent \}\\
\\
\indent if(k[p] == 1)\\
\indent \indent k <- k[-p]\\
\\
\indent do.demean <- rep(T, length(k))\\
\indent do.demean[1] <- F\\
\\
\indent if(top.down)\\
\indent \{\\
\indent \indent k1 <- c(k[-1], 1)\\
\indent \indent k <- round(k / k1)[length(k):1]\\
\indent \indent ind.list <- qrm.stat.class(ret, k,\\
\indent \indent \indent iter.max = iter.max, num.try = num.try)\\
\indent \}\\
\indent else\\
\indent \indent ind.list <- qrm.stat.ind.class.all(ret, k,\\
\indent \indent \indent iter.max = iter.max, num.try = num.try, do.demean = do.demean)\\
\\
\indent return(ind.list)\\
\}\\
\\
\noindent qrm.eigen <- function (ret, calc.cor = F)\\
\{\\
\indent if(ncol(ret) - 1 <= nrow(ret))\\
\indent \indent return(qrm.calc.eigen.eff(ret, calc.cor = calc.cor))\\
\\
\indent return(eigen(qrm.calc.cov.mat(ret, calc.cor = calc.cor)))\\
\}\\
\\
\noindent qrm.calc.cov.mat <- function(x, calc.cor = F)\\
\{\\
\indent tr <- apply(x, 1, sd)\\
\indent x <- x / tr\\
\indent x <- x - rowMeans(x)\\
\indent y <- x \%*\% t(x) / (ncol(x) - 1)\\
\indent return(y)\\
\}
}}

\subsection{Code for Hybrid Industry Classification}

{}In this subsection we give the R source code for hybrid industry classifications discussed in Section \ref{sec.hybrid}. There is only one function {\tt{\small{qrm.improve.ind.class(ret, ind, iter.max = 10, num.try = 100)}}}, which internally calls the main function from Subsection \ref{app.A.main} {\tt{\small{qrm.stat.ind.class()}}} with the same inputs {\tt{\small{ret}}}, {\tt{\small{iter.max}}} and {\tt{\small{num.try}}}, and the following new input: {\tt{\small{ind}}} is an $N\times K_*$ binary industry classification matrix corresponding to the most granular level of a ``fundamental" industry classification (e.g., sub-industries in BICS). The output is an $N\times K_*^\prime$ binary industry classification matrix {\tt{\small{ind1}}}. Here $K^\prime_* \geq K_*$. Typically $K^\prime_* > K_*$, so we get a more granular industry classification after clustering. If $K^\prime_* = K_*$, then {\tt{\small{ind1}}} is the same as {\tt{\small{ind}}}.\\
\\
{{\tt\small
\noindent qrm.improve.ind.class <- function (ret, ind,\\
\indent iter.max = 10, num.try = 100)\\
\{\\
\indent ind1 <- rep(NA, nrow(ret))\\
\indent for(i in 1:ncol(ind))\\
\indent \{\\
\indent \indent k <- round(sum(ind[, i]) / (ncol(ret)-1))\\
\indent \indent if(k < 2)\\
\indent \indent \{\\
\indent \indent \indent ind1 <- cbind(ind1, ind[, i])\\
\indent \indent \indent next\\
\indent \indent \}\\
\indent \indent take <- ind[, i] > 0\\
\indent \indent x <- ret[take, ]\\
\indent \indent y <- qrm.stat.ind.class(x, k,\\
\indent \indent \indent iter.max = iter.max, num.try = num.try)\\
\indent \indent if(length(y) > sum(take))\\
\indent \indent \indent tmp <- matrix(0, nrow(ret), ncol(y))\\
\indent \indent else\\
\indent \indent \indent tmp <- matrix(0, nrow(ret), 1)\\
\\
\indent \indent tmp[take, ] <- y\\
\indent \indent ind1 <- cbind(ind1, tmp)\\
\indent \}\\
\indent ind1 <- ind1[, -1]\\
\indent return(ind1)\\
\}
}}

\section{R Code for Top-Down Clustering}\label{app.B}

{}In this Appendix we give the R source code for building multilevel ``top-down" statistical industry classifications (see Subsection \ref{sub.topdown}): {\tt{\small{qrm.stat.class(ret, k, iter.max = 10, num.try = 100)}}} internally calls
{\tt{\small{qrm.stat.ind.class()}}} defined in Subsection \ref{app.A.main} with the same inputs {\tt{\small{ret}}}, {\tt{\small{iter.max}}} and {\tt{\small{num.try}}}, and the following new input: {\tt{\small{k}}} $= (L_P, L_{P-1}, \dots, L_2, L_1)$ is a {\em reversed} $P$-vector $L_\mu$, $\mu=1,\dots,P$, defined in Subsection \ref{sub.topdown}, and $P$ is the number of levels. The output is a list {\tt{\small{ind.list}}} with $P$ members, same as in Subsection \ref{app.A.all}.\\
\\
{\tt\small{
\noindent qrm.stat.class <- function (ret, k, iter.max = 10, num.try = 100)\\
\{\\
\indent k <- c(1, k)\\
\indent n <- nrow(ret)\\
\indent p <- length(k)\\
\indent ind <- list()\\
\indent ind.list <- list()\\
\indent for(lvl in 1:p)\\
\indent \indent ind[[lvl]] <- matrix(1, n, 1)\\
\\
\indent for(lvl in 2:p)\\
\indent \{\\
\indent \indent for(a in 1:ncol(ind[[lvl - 1]]))\\
\indent \indent \{\\
\indent \indent \indent take <- ind[[lvl - 1]][, a] > 0\\
\indent \indent \indent tmp.k <- sum(take)\\
\indent \indent \indent if(tmp.k <= k[lvl])\\
\indent \indent \indent \{\\
\indent \indent \indent \indent ind[[lvl]] <- cbind(ind[[lvl]], as.numeric(take))\\
\indent \indent \indent \indent next\\
\indent \indent \indent \}\\
\indent \indent \indent x <- matrix(ret[take, ], tmp.k, ncol(ret))\\
\indent \indent \indent norm.x <- qrm.calc.norm.ret(x)\\
\indent \indent \indent tmp.ind <- qrm.stat.ind.class(x, k[lvl],\\
\indent \indent \indent \indent iter.max, num.try = num.try)\\
\indent \indent \indent if(length(tmp.ind) > tmp.k)\\
\indent \indent \indent \indent tmp <- matrix(0, n, ncol(tmp.ind))\\
\indent \indent \indent else\\
\indent \indent \indent \indent tmp <- matrix(0, n, 1)\\
\\
\indent \indent \indent tmp[take, ] <- tmp.ind\\
\indent \indent \indent ind[[lvl]] <- cbind(ind[[lvl]], tmp)\\
\indent \indent \}\\
\indent \indent ind[[lvl]] <- ind[[lvl]][, -1]\\
\indent \}\\
\\
\indent for(lvl in p:2)\\
\indent \indent ind.list[[p - lvl + 1]] <- ind[[lvl]]\\
\\
\indent return(ind.list)\\
\}
}}

\section{R Code for Relaxation Clustering}\label{app.C}

{}{}In this Appendix we give the R source code for building relaxation algorithm based multilevel statistical industry classifications (see Subsection \ref{relax}). The first function, {\tt{\small{qrm.stat.clust.all(ret, k, do.demean = rep(F, length(k)), norm.cl.ret = F)}}},
is essentially the same as the function {\tt{\small{qrm.stat.ind.class.all()}}} in Subsection \ref{app.A.all}, except internally it calls the within function {\tt{\small{qrm.stat.clust(ret, k, demean.ret = F, return.clust = F)}}}. The latter builds a relaxation based single-level classification with {\tt{\small{k}}} clusters. The additional input is {\tt{\small{return.clust}}}: when set to {\tt{\small{TRUE}}}, this function outputs the $N$-vector $G(i)$ as opposed to the $N\times K$ binary industry classification matrix (as for the default value). Recall that $G:\{1,\dots,N\}\mapsto\{1,\dots,K\}$ maps stocks to clusters.\\
\\
{\tt{\small
\noindent qrm.stat.clust.all <- function (ret, k,\\
\indent do.demean = rep(F, length(k)), norm.cl.ret = F)\\
\{\\
\indent ind.list <- list()\\
\\
\indent for(i in 1:length(k))\\
\indent \{\\
\indent \indent ind.list[[i]] <- qrm.stat.clust(ret, k[i],\\
\indent \indent \indent demean.ret = do.demean[i])\\
\indent \indent if(norm.cl.ret)\\
\indent \indent \indent ret <- t(ind.list[[i]]) \%*\% qrm.calc.norm.ret(ret)\\
\indent \indent else\\
\indent \indent \indent ret <- t(ind.list[[i]]) \%*\% ret\\
\\
\indent \indent if(i > 1)\\
\indent \indent \{\\
\indent \indent \indent ind.list[[i]] <- ind.list[[i - 1]] \%*\% ind.list[[i]]\\
\indent \indent \indent take <- ind.list[[i]] > 0\\
\indent \indent \indent ind.list[[i]][take] <- 1\\
\indent \indent \}\\
\indent \}\\
\\
\indent return(ind.list)\\
\}\\
\\
\noindent qrm.stat.clust <- function (ret, k,\\
\indent demean.ret = F, return.clust = F)\\
\{\\
\indent calc.take <- function(n, ix, q)\\
\indent \{\\
\indent \indent q1 <- q[ix > q]\\
\indent \indent q2 <- q[ix < q]\\
\indent \indent take1 <- ix + (q1 - 1) * n\\
\indent \indent take2 <- q2 + (ix - 1) * n\\
\indent \indent take <- c(take1, take2)\\
\indent \indent return(take)\\
\indent \}\\
\\
\indent calc.dist.mat <- function(x)\\
\indent \{\\
\indent \indent if(is.matrix(x))\\
\indent \indent \indent n <- nrow(x)\\
\indent \indent else\\
\indent \indent \indent n <- length(x)\\
\\
\indent \indent y <- x \%*\% t(x)\\
\indent \indent z <- matrix(diag(y), n, n)\\
\indent \indent y <- z + t(z) - 2 * y\\
\indent \indent take <- upper.tri(y, T)\\
\indent \indent y[take] <- NA\\
\indent \indent return(y)\\
\indent \}\\
\\
\indent extract.ix <- function(y)\\
\indent \{\\
\indent \indent k <- as.numeric(y[1])\\
\indent \indent j <- trunc(k / n)\\
\indent \indent if(j == k / n)\\
\indent \indent \indent i <- n\\
\indent \indent else\\
\indent \indent \{\\
\indent \indent \indent i <- k - j * n\\
\indent \indent \indent j <- j + 1\\
\indent \indent \}\\
\indent \indent return(c(i, j))\\
\indent \}\\
\\
\indent if(demean.ret)\\
\indent \indent ret <- t(t(ret) - colMeans(ret))\\
\\
\indent n <- nrow(ret)\\
\indent v <- 1:n\\
\\
\indent ret <- qrm.calc.norm.ret(ret)\\
\indent x <- calc.dist.mat(ret)\\
\indent x <- as.vector(x)\\
\indent names(x) <- as.character(1:length(x))\\
\indent x <- sort(x)\\
\indent y <- as.numeric(names(x))\\
\\
\indent m <- 0\\
\indent count <- 0\\
\indent w <- rep(0, n)\\
\indent set.y <- F\\
\\
\indent while(count < n)\\
\indent \{\\
\indent \indent if(m < k)\\
\indent \indent \indent y1 <- y\\
\indent \indent else if(!set.y)\\
\indent \indent \{\\
\indent \indent \indent set.y <- T\\
\indent \indent \indent q <- v[w == 0]\\
\indent \indent \indent n1 <- length(q)\\
\indent \indent \indent u <- q + matrix((q - 1) * n, n1, n1, byrow = T)\\
\indent \indent \indent take <- upper.tri(u, T)\\
\indent \indent \indent u <- as.vector(u[!take])\\
\indent \indent \indent take <- !(y \%in\% u)\\
\indent \indent \indent y1 <- y[take]\\
\indent \indent \}\\
\indent \indent else\\
\indent \indent \{\\
\indent \indent \indent q <- v[w == 0]\\
\indent \indent \indent take <- calc.take(n, p, q)\\
\indent \indent \indent take <- !(u \%in\% take)\\
\indent \indent \indent u <- u[take]\\
\indent \indent \indent take <- !(y \%in\% u)\\
\indent \indent \indent y1 <- y[take]\\
\indent \indent \}\\
\\
\indent \indent ix <- extract.ix(y1)\\
\indent \indent q <- v[w > 0]\\
\\
\indent \indent if(w[ix[1]] > 0)\\
\indent \indent \{\\
\indent \indent \indent count <- count + 1\\
\indent \indent \indent w[p <- ix[2]] <- w[ix[1]]\\
\indent \indent \indent take <- calc.take(n, p, q)\\
\indent \indent \}\\
\indent \indent else if(w[ix[2]] > 0)\\
\indent \indent \{\\
\indent \indent \indent count <- count + 1\\
\indent \indent \indent w[p <- ix[1]] <- w[ix[2]]\\
\indent \indent \indent take <- calc.take(n, p, q)\\
\indent \indent \}\\
\indent \indent else\\
\indent \indent \{\\
\indent \indent \indent m <- m + 1\\
\indent \indent \indent count <- count + 2\\
\indent \indent \indent w[ix] <- m\\
\indent \indent \indent take <- c(calc.take(n, ix[1], q), calc.take(n, ix[2], q))\\
\indent \indent \}\\
\\
\indent \indent take <- c(take, ix[1] + (ix[2] - 1) * n)\\
\indent \indent take <- !(y \%in\% take)\\
\indent \indent y <- y[take]\\
\indent \}\\
\\
\indent if(return.clust)\\
\indent \indent return(w)\\
\\
\indent k <- min(k, m)\\
\indent z <- matrix(NA, n, k)\\
\indent for(j in 1:k)\\
\indent \indent z[, j] <- as.numeric(w == j)\\
\\
\indent return(z)\\
\}
}}

\section{DISCLAIMERS}\label{app.D}

{}Wherever the context so requires, the masculine gender includes the feminine and/or neuter, and the singular form includes the plural and {\em vice versa}. The author of this paper (``Author") and his affiliates including without limitation Quantigic$^\circledR$ Solutions LLC (``Author's Affiliates" or ``his Affiliates") make no implied or express warranties or any other representations whatsoever, including without limitation implied warranties of merchantability and fitness for a particular purpose, in connection with or with regard to the content of this paper including without limitation any code or algorithms contained herein (``Content").

{}The reader may use the Content solely at his/her/its own risk and the reader shall have no claims whatsoever against the Author or his Affiliates and the Author and his Affiliates shall have no liability whatsoever to the reader or any third party whatsoever for any loss, expense, opportunity cost, damages or any other adverse effects whatsoever relating to or arising from the use of the Content by the reader including without any limitation whatsoever: any direct, indirect, incidental, special, consequential or any other damages incurred by the reader, however caused and under any theory of liability; any loss of profit (whether incurred directly or indirectly), any loss of goodwill or reputation, any loss of data suffered, cost of procurement of substitute goods or services, or any other tangible or intangible loss; any reliance placed by the reader on the completeness, accuracy or existence of the Content or any other effect of using the Content; and any and all other adversities or negative effects the reader might encounter in using the Content irrespective of whether the Author or his Affiliates is or are or should have been aware of such adversities or negative effects.

{}The R code included in Appendix \ref{app.A}, Appendix \ref{app.B} and Appendix \ref{app.C} hereof is part of the copyrighted R code of Quantigic$^\circledR$ Solutions LLC and is provided herein with the express permission of Quantigic$^\circledR$ Solutions LLC. The copyright owner retains all rights, title and interest in and to its copyrighted source code included in Appendix \ref{app.A}, Appendix \ref{app.B} and Appendix \ref{app.C} hereof and any and all copyrights therefor.

\newpage

\begin{table}[ht]
\caption{Simulation results (11 runs) for the optimized alphas with bounds using heterotic risk models based on ``bottom-up" statistical industry classifications obtained via a single sampling in each run. The numbers of clusters in a 3-level hierarchy are 100, 30 and 10. See Subsection \ref{sub.bottomup} and Section \ref{sec.backtests} for details.} 
\begin{tabular}{l l l l} 
\hline\hline 
Run & ROC & SR & CPS\\[0.5ex] 
\hline 
1 & 41.396\% & 16.195 & 2.060\\
2 & 41.572\% & 16.091 & 2.065\\
3 & 41.666\% & 16.318 & 2.070\\
4 & 41.544\% & 16.300 & 2.065\\
5 & 41.455\% & 16.238 & 2.058\\
6 & 41.731\% & 16.251 & 2.074\\
7 & 41.391\% & 16.238 & 2.057\\
8 & 41.567\% & 16.293 & 2.065\\
9 & 41.755\% & 16.135 & 2.075\\
10 & 41.627\% & 16.122 & 2.068\\
11 & 41.569\% & 16.260 & 2.065\\[1ex] 
\hline 
\end{tabular}
\label{table.try1.100.30.10} 
\end{table}

\begin{table}[ht]
\caption{Simulation results (11 runs) for the optimized alphas with bounds using heterotic risk models based on ``bottom-up" statistical industry classifications obtained via aggregating 100 samplings in each run. The target number of clusters for a single level is 100. See Subsection \ref{sub.aggr} and Section \ref{sec.backtests} for details.} 
\begin{tabular}{l l l l} 
\hline\hline 
Run & ROC & SR & CPS\\[0.5ex] 
\hline 
1 & 41.907\% & 16.427 & 2.103\\
2 & 41.912\% & 16.210 & 2.100\\
3 & 41.774\% & 16.227 & 2.091\\
4 & 41.811\% & 16.295 & 2.094\\
5 & 41.832\% & 16.263 & 2.092\\
6 & 42.047\% & 16.102 & 2.109\\
7 & 41.839\% & 16.242 & 2.098\\
8 & 41.966\% & 16.027 & 2.104\\
9 & 41.841\% & 15.941 & 2.096\\
10 & 41.755\% & 16.131 & 2.093\\
11 & 41.775\% & 16.284 & 2.093\\[1ex] 
\hline 
\end{tabular}
\label{table.try100.100} 
\end{table}

\begin{table}[ht]
\caption{Simulation results (23 runs) for the optimized alphas with bounds using heterotic risk models based on statistical industry classifications obtained via aggregating 100 samplings in each run. The target numbers of clusters in a 3-level hierarchy are 100, 30 and 10. See Subsection \ref{sub.aggr} and Section \ref{sec.backtests} for details. The first 15 runs correspond to {\tt{\small{norm.cl.ret = F}}}, the other 8 runs correspond to {\tt{\small{norm.cl.ret = T}}}; see the function {\tt{\small{qrm.stat.ind.class.all()}}} in Appendix \ref{app.A}.} 
\begin{tabular}{l l l l} 
\hline\hline 
Run & ROC & SR & CPS\\[0.5ex] 
\hline 
1 & 42.181\% & 16.565 & 2.113\\
2 & 41.728\% & 16.314 & 2.092\\
3 & 41.895\% & 16.419 & 2.097\\
4 & 41.958\% & 16.350 & 2.103\\
5 & 42.034\% & 16.373 & 2.106\\
6 & 41.700\% & 16.149 & 2.093\\
7 & 42.134\% & 16.055 & 2.112\\
8 & 42.113\% & 16.150 & 2.109\\
9 & 41.586\% & 16.288 & 2.083\\
10 & 41.808\% & 16.267 & 2.094\\
11 & 41.925\% & 16.168 & 2.099\\
12 & 41.861\% & 16.228 & 2.096\\
13 & 41.766\% & 16.223 & 2.093\\
14 & 41.877\% & 16.331 & 2.095\\
15 & 42.148\% & 16.217 & 2.112\\
16 & 41.895\% & 16.240 & 2.099\\
17 & 41.857\% & 16.252 & 2.099\\
18 & 41.777\% & 16.169 & 2.092\\
19 & 41.886\% & 16.341 & 2.101\\
20 & 41.851\% & 16.207 & 2.094\\
21 & 42.266\% & 16.144 & 2.119\\
22 & 41.769\% & 16.205 & 2.093\\
23 & 42.083\% & 16.095 & 2.110\\[1ex] 
\hline 
\end{tabular}
\label{table.try100.100.30.10} 
\end{table}

\begin{table}[ht]
\noindent
\caption{Summaries of the actual numbers of clusters in a statistical industry classification obtained via aggregating 100 samplings. The target numbers of clusters in a 3-level hierarchy are 100, 30 and 10. The summaries are based on 60 data points corresponding to sixty 21-trading-day intervals in the 1,260 trading-day backtesting period. See Subsection \ref{sub.aggr} and Section \ref{sec.backtests} for details. 1st Qu. = 1st Quartile, 3rd Qu. = 3rd Quartile, StDev = standard deviation, MAD = mean absolute deviation. The 100 samplings correspond to the run reported in the last row of Table \ref{table.try100.100.30.10}.}
\begin{tabular}{l l l l l l l l l} 
\\
\hline\hline 
Level & Min & 1st Qu. & Median & Mean & 3rd Qu. & Max & StDev & MAD \\[0.5ex] 
\hline 
1 & 87 &  93 &  94 & 93.95 &   96 &  99 & 2.33 & 1.48\\
2 & 20 &  24 &  25 & 24.93 &   26 &  28 & 1.91 & 1.48\\
3 &  6 &   8 &   9 &  8.58 &    9 &  10 & 0.93 & 1.48\\ [1ex] 
\hline 
\end{tabular}
\label{table.num.clust} 
\end{table}

\begin{table}[ht]
\caption{Simulation results for the optimized alphas with bounds using heterotic risk models based on ``top-down" 3-level statistical industry classifications obtained via a single sampling in each run, with 3 runs for each choice of the 3-vector ${\widetilde L}_\mu = (L_3, L_2, L_1)$, which is the {\em reverse} of the 3-vector $L_\mu$, $\mu=1,2,3$, defined in Subsection \ref{sub.topdown}. Also see Section \ref{sec.backtests} for details.} 
\begin{tabular}{l l l l l} 
\hline\hline 
\noalign{\vskip 1mm}
Run & ${\widetilde L}_\mu$ & ROC & SR & CPS\\[0.5ex] 
\hline 
1 & (10,5,5) & 40.637\% & 16.502 & 2.055\\
2 & (10,5,5) & 40.880\% & 16.511 & 2.070\\
3 & (10,5,5) & 40.902\% & 16.684 & 2.075\\
1 & (10,5,3) & 41.278\% & 16.274 & 2.077\\
2 & (10,5,3) & 41.274\% & 16.342 & 2.076\\
3 & (10,5,3) & 41.044\% & 16.248 & 2.063\\
1 & (10,3,4) & 41.334\% & 16.046 & 2.071\\
2 & (10,3,4) & 41.442\% & 16.236 & 2.077\\
3 & (10,3,4) & 41.343\% & 16.130 & 2.071\\
1 & (10,3,3) & 41.590\% & 16.102 & 2.074\\
2 & (10,3,3) & 41.620\% & 16.029 & 2.072\\
3 & (10,3,3) & 41.654\% & 16.048 & 2.076\\
1 & (10,2,3) & 41.553\% & 15.724 & 2.054\\
2 & (10,2,3) & 42.046\% & 16.027 & 2.081\\
3 & (10,2,3) & 41.765\% & 15.665 & 2.066\\
1 & (10,2,2) & 42.144\% & 15.598 & 2.076\\
2 & (10,2,2) & 41.925\% & 15.516 & 2.063\\
3 & (10,2,2) & 42.007\% & 15.553 & 2.066\\[1ex] 
\hline 
\end{tabular}
\label{table.try1.topdown} 
\end{table}

\begin{table}[ht]
\caption{Simulation results for the optimized alphas with bounds using heterotic risk models based on ``top-down" 3-level statistical industry classifications obtained via aggregating 100 samplings in each run, with 3 runs for each ${\widetilde L}_\mu = (L_3, L_2, L_1)$, which is the {\em reverse} of the 3-vector $L_\mu$, $\mu=1,2,3$, defined in Subsection \ref{sub.topdown}. Also see Section \ref{sec.backtests} for details.} 
\begin{tabular}{l l l l l} 
\hline\hline 
\noalign{\vskip 1mm}
Run & ${\widetilde L}_\mu$ & ROC & SR & CPS\\[0.5ex] 
\hline 
1 & (10,5,5) & 41.412\% & 16.550 & 2.098\\
2 & (10,5,5) & 41.478\% & 16.413 & 2.097\\
3 & (10,5,5) & 41.251\% & 16.401 & 2.092\\
1 & (10,5,3) & 41.696\% & 16.057 & 2.095\\
2 & (10,5,3) & 41.597\% & 16.157 & 2.093\\
3 & (10,5,3) & 41.730\% & 15.975 & 2.100\\
1 & (10,3,4) & 41.680\% & 15.979 & 2.085\\
2 & (10,3,4) & 41.643\% & 15.903 & 2.078\\
3 & (10,3,4) & 41.794\% & 16.023 & 2.092\\
1 & (10,3,3) & 42.078\% & 15.975 & 2.090\\
2 & (10,3,3) & 41.897\% & 15.962 & 2.083\\
3 & (10,3,3) & 41.785\% & 15.904 & 2.078\\
1 & (10,2,3) & 41.817\% & 15.618 & 2.063\\
2 & (10,2,3) & 41.964\% & 15.693 & 2.071\\
3 & (10,2,3) & 41.705\% & 15.598 & 2.062\\
1 & (10,2,2) & 42.080\% & 15.489 & 2.065\\
2 & (10,2,2) & 41.865\% & 15.433 & 2.059\\
3 & (10,2,2) & 41.987\% & 15.468 & 2.063\\[1ex] 
\hline 
\end{tabular}
\label{table.try100.topdown} 
\end{table}

\begin{table}[ht]
\caption{Simulation results for the optimized alphas with bounds using heterotic risk models based on ``bottom-up" $P$-level statistical industry classifications obtained via aggregating 100 samplings in each run, with multiple (3 or 4) runs for each $P$. The cluster numbers $K_\mu$, $\mu=1,\dots,P$, are determined dynamically via the algorithm of Subsection \ref{sub.dyn}. Also see Section \ref{sec.backtests} for backtesting details.} 
\begin{tabular}{l l l l l} 
\hline\hline 
Run & $P$ & ROC & SR & CPS\\[0.5ex] 
\hline 
1 & 2 & 41.746\% & 16.152 & 2.093\\
2 & 2 & 41.745\% & 16.004 & 2.091\\
3 & 2 & 42.029\% & 16.007 & 2.104\\
1 & 3 & 41.921\% & 16.309 & 2.103\\
2 & 3 & 41.911\% & 16.090 & 2.098\\
3 & 3 & 41.813\% & 16.455 & 2.094\\
1 & 4 & 41.887\% & 16.317 & 2.096\\
2 & 4 & 42.273\% & 16.168 & 2.117\\
3 & 4 & 41.850\% & 16.115 & 2.099\\
1 & 5 & 42.095\% & 16.359 & 2.112\\
2 & 5 & 41.891\% & 16.178 & 2.102\\
3 & 5 & 41.961\% & 16.278 & 2.101\\
4 & 5 & 42.152\% & 16.237 & 2.111\\[1ex] 
\hline 
\end{tabular}
\label{table.dyn} 
\end{table}

\begin{table}[ht]
\caption{Simulation results for the optimized alphas with bounds using heterotic risk models based on ``bottom-up" statistical industry classifications obtained via aggregating 100 samplings in each run. $K$ is the target number of clusters for a single level. The $K=100$ entry is the same as the last row in Table \ref{table.try100.100}. See Subsection \ref{sub.aggr} and Section \ref{sec.backtests} for details. Also see Figures 1, 2 and 3.} 
\begin{tabular}{l l l l} 
\hline\hline 
$K$ & ROC & SR & CPS\\[0.5ex] 
\hline 
10 & 41.726\% & 14.853 & 2.027\\
25 & 42.024\% & 15.395 & 2.065\\
50 & 42.180\% & 15.941 & 2.094\\
75 & 41.771\% & 16.115 & 2.085\\
100 & 41.775\% & 16.284 & 2.093\\
125 & 41.427\% & 16.205 & 2.080\\
150 & 41.306\% & 16.337 & 2.073\\
175 & 41.286\% & 16.456 & 2.076\\
200 & 40.774\% & 16.276 & 2.047\\
250 & 40.611\% & 16.248 & 2.032\\[1ex] 
\hline 
\end{tabular}
\label{table.try100.K} 
\end{table}

\begin{table}[ht]
\caption{Summary of stock counts (first column) for the 10 (out of 165 in this sample) most populous BICS sub-industries (most granular level, second column) for 2000 stocks in our backtests for a randomly chose date. We also show the corresponding BICS industries (less granular level, third column) and  BICS sectors (least granular level, fourth column). The nomenclature is shown as it appears in BICS.} 
\begin{tabular}{l l l l} 
\hline\hline 
\#(stocks) & BICS Sub-industry & BICS Industry & BICS Sector\\[0.5ex] 
\hline 
94 & Banks & Banking & Financials\\
94 & REIT & Real Estate & Financials\\
74 & Exploration \& Production & Oil, Gas \& Coal & Energy\\
52 & Semiconductor Devices & Semiconductors & Technology\\
50 & Application Software & Software & Technology\\
47 & Utility Networks & Utilities & Utilities\\
46 & Telecom Carriers & Telecom & Communications\\
45 & Oil \& Gas Services \& Equip & Oil, Gas \& Coal & Energy\\
44 & Biotech & Biotech \& Pharma & Health Care\\
38 & Specialty Pharma & Biotech \& Pharma & Health Care\\[1ex] 
\hline 
\end{tabular}
\label{table.bics.summary} 
\end{table}

\begin{table}[ht]
\caption{Simulation results (14 runs) for the optimized alphas with bounds using heterotic risk models based on hybrid industry classifications (see Section \ref{sec.hybrid}) using statistical industry classifications based on aggregating 100 samplings in each run.} 
\begin{tabular}{l l l l} 
\hline\hline 
Run & ROC & SR & CPS\\[0.5ex] 
\hline 
1 & 49.214\% & 19.447 & 2.380\\
2 & 49.260\% & 19.571 & 2.387\\
3 & 49.224\% & 19.528 & 2.386\\
4 & 49.126\% & 19.522 & 2.379\\
5 & 49.217\% & 19.506 & 2.384\\
6 & 49.163\% & 19.547 & 2.382\\
7 & 49.204\% & 19.517 & 2.384\\
8 & 49.138\% & 19.482 & 2.381\\
9 & 49.247\% & 19.529 & 2.385\\
10 & 49.195\% & 19.504 & 2.384\\
11 & 49.191\% & 19.550 & 2.383\\
12 & 49.216\% & 19.578 & 2.383\\
13 & 49.212\% & 19.519 & 2.385\\
14 & 49.307\% & 19.537 & 2.389\\[1ex] 
\hline 
\end{tabular}
\label{table.improve} 
\end{table}

\begin{table}[ht]
\noindent
\caption{Summaries of the numbers of top 10 most populous: (i) BICS sub-industries before clustering (first 10 rows); and (ii) resultant clusters at the same level in a hybrid industry classification after clustering (last 10 rows). Each summary is over 60 datapoints   (see Section \ref{sec.hybrid}).}
\begin{tabular}{l l l l l l l l l} 
\\
\hline\hline 
Order & Min & 1st Qu. & Median & Mean & 3rd Qu. & Max & StDev & MAD \\[0.5ex] 
\hline 
1 & 89 & 93 & 94 & 93.8 & 95 & 98 & 1.964 & 1.483\\
2 & 81 & 89 & 91 & 90.37 & 92 & 96 & 3.103 & 2.965\\
3 & 63 & 69 & 73 & 72.28 & 75 & 81 & 4.388 & 4.448\\
4 & 50 & 54 & 56 & 56.5 & 59 & 65 & 3.327 & 3.706\\
5 & 49 & 50 & 51 & 51.65 & 53 & 56 & 1.745 & 1.483\\
6 & 46 & 49 & 49 & 49.18 & 50 & 51 & 1.255 & 1.483\\
7 & 44 & 45.75 & 47 & 47.22 & 49 & 50 & 1.869 & 2.965\\
8 & 41 & 44 & 45.5 & 45.58 & 47 & 50 & 1.977 & 2.224\\
9 & 36 & 38.75 & 40 & 40.18 & 41 & 46 & 2.318 & 1.483\\
10 & 34 & 37 & 37 & 37.77 & 39 & 45 & 1.925 & 1.483\\
1 & 44 & 51.75 & 57 & 58.82 & 64 & 85 & 8.981 & 8.896\\
2 & 33 & 45 & 49 & 49 & 53.25 & 69 & 6.857 & 5.93\\
3 & 32 & 39.75 & 44 & 43.28 & 46.25 & 56 & 4.854 & 4.448\\
4 & 31 & 37 & 40 & 40.22 & 44 & 47 & 4.030 & 4.448\\
5 & 31 & 36 & 37 & 37.85 & 40.25 & 46 & 3.473 & 2.965\\
6 & 29 & 33.75 & 35 & 35.42 & 37 & 45 & 3.285 & 2.965\\
7 & 29 & 31.75 & 34 & 33.78 & 36 & 41 & 2.964 & 2.965\\
8 & 28 & 30 & 32 & 31.85 & 34 & 38 & 2.543 & 2.965\\
9 & 26 & 29 & 30.5 & 30.67 & 32.25 & 35 & 2.252 & 2.224\\
10 & 25 & 28 & 29 & 29.2 & 31 & 35 & 2.122 & 2.965\\ [1ex] 
\hline 
\end{tabular}
\label{table.ind.counts} 
\end{table}
\clearpage
\newpage
\begin{figure}[ht]
\centerline{\epsfxsize 4.truein \epsfysize 4.truein\epsfbox{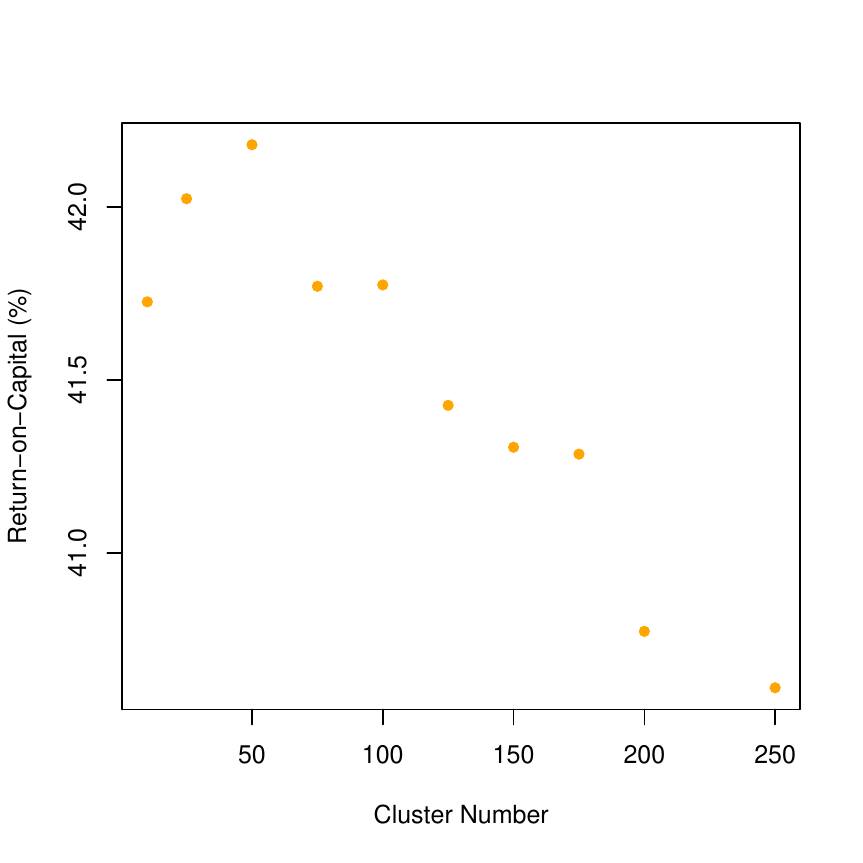}}
\noindent{\small {Figure 1. Graph of the values of the return-on-capital (ROC) in percent from Table \ref{table.try100.K} vs. the target number of clusters $K$ (as defined in said table).}}
\end{figure}

\clearpage
\newpage
\begin{figure}[ht]
\centerline{\epsfxsize 4.truein \epsfysize 4.truein\epsfbox{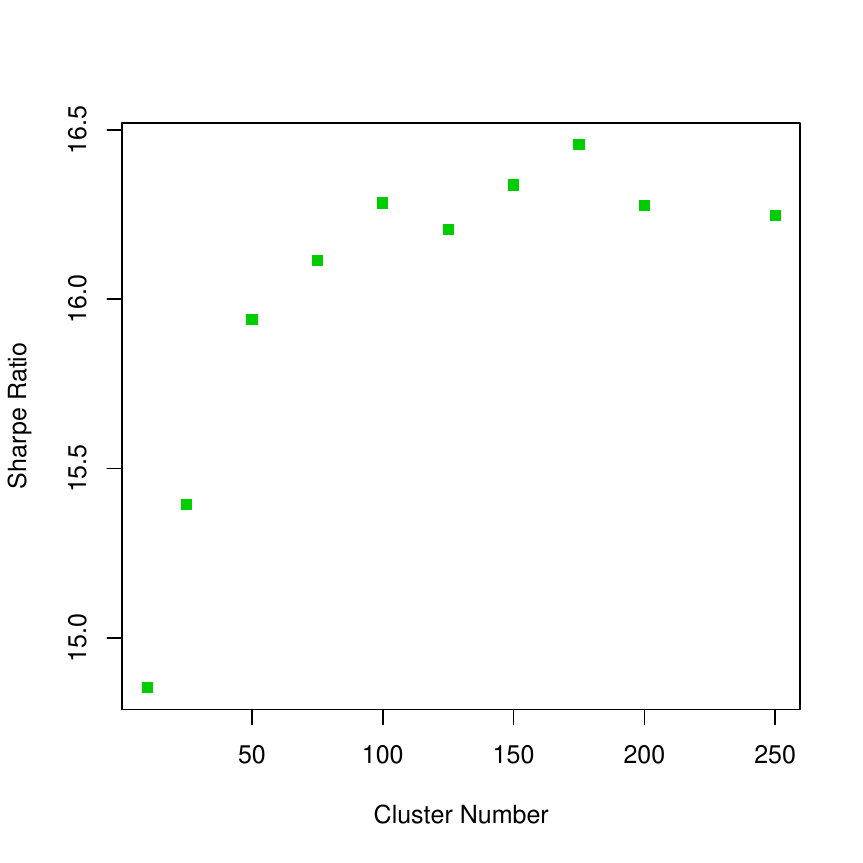}}
\noindent{\small {Figure 2. Graph of the values of the Sharpe ratio (SR) from Table \ref{table.try100.K} vs. the target number of clusters $K$ (as defined in said table).}}
\end{figure}

\clearpage
\newpage
\begin{figure}[ht]
\centerline{\epsfxsize 4.truein \epsfysize 4.truein\epsfbox{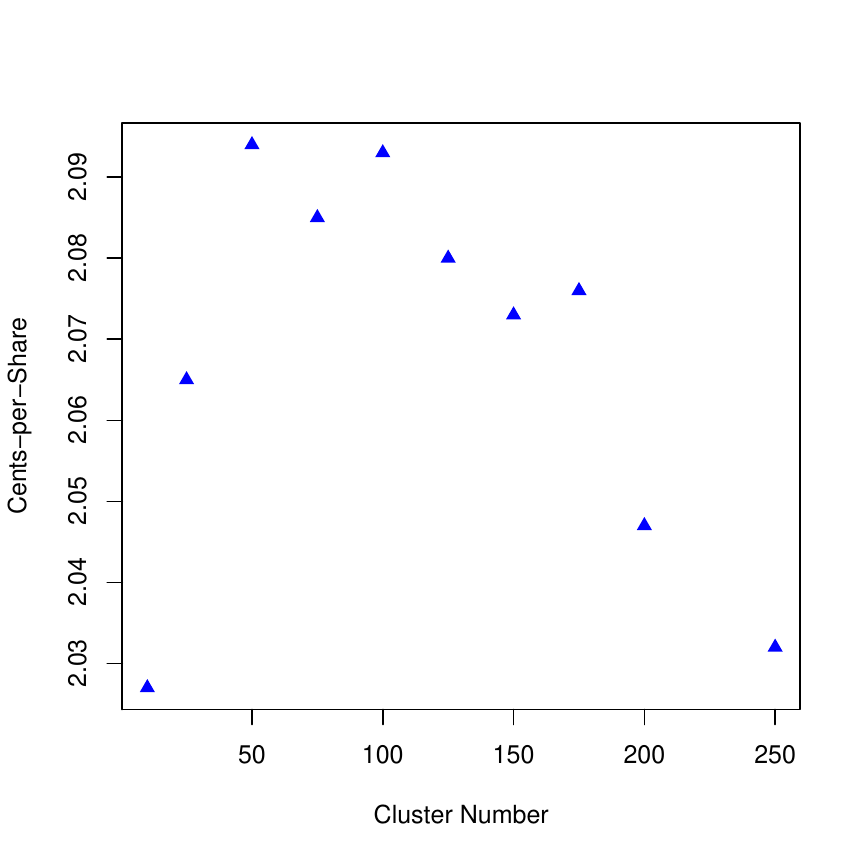}}
\noindent{\small {Figure 3. Graph of the values of cents-per-share from Table \ref{table.try100.K} vs. the target number of clusters $K$ (as defined in said table).}}
\end{figure}

\end{document}